\documentclass[final, 5p, times]{elsarticle}
\usepackage[T1]{fontenc}
\usepackage[latin9]{inputenc}
\usepackage{amsmath}
\usepackage{amssymb}
\usepackage{graphicx}
\usepackage{esint}
\usepackage[caption=false, font=footnotesize]{subfig}

\usepackage{booktabs}

\journal{NIMA}
\usepackage{siunitx}[=v2]
\usepackage{upgreek}
\usepackage[bookmarks,bookmarksopen,bookmarksdepth=6]{hyperref}
\hypersetup{
  colorlinks=true,
  linkcolor=blue,
  filecolor=magenta,      
  urlcolor=cyan
}

\usepackage{cases}
\usepackage{url}
\usepackage{lineno}
\DeclareSIUnit\particles{n}
\DeclareSIQualifier\eq{eq}

\makeatother

\begin{document}
\title{Evaluation of HPK $n^+$-$p$ planar pixel sensors for
the CMS Phase-2 upgrade}

\author[]{The Tracker Group of the CMS Collaboration\corref{cor1}\fnref{fn1}}
\cortext[cor1]{Corresponding author, Email address: Joern.Schwandt@desy.de, Tel.: +49 40 8998 4742}
\fntext[fn1]{Complete author list at the end of document.}

\begin{abstract}
To cope with the challenging environment of the planned high luminosity upgrade
of the Large Hadron Collider (HL-LHC), scheduled to start operation in 2029,
CMS will replace its entire tracking system.
The requirements for the tracker are largely 
determined by the long operation time of 10~years
with an instantaneous peak luminosity of up to 
\SI{7.5e34}{\per\square\centi\meter\per\second} in the ultimate 
performance scenario.
Depending on the radial distance from the interaction point, the silicon sensors will receive a particle fluence corresponding to a non-ionizing energy loss of up to $\Phi_{\text{eq}} = $ \SI{3.5e16}{\per\centi\meter\squared}. 
This paper focuses on planar pixel sensor design and qualification up to a fluence of $\Phi_{\text{eq}} = \SI{1.4e16}{\per\square\centi\meter}$.


For the development of appropriate planar pixel 
sensors an R\&D program was initiated, which includes $n^+$-$p$ sensors on 
\SI{150}{\milli\meter} (6'')
wafers with an active thickness of \SI{150}{\micro\meter} with pixel sizes of 
\num{100x25}~\si{\square\micro\meter} and 
\num{50x50}~\si{\square\micro\meter} manufactured by Hamamatsu. Single chip 
modules with ROC4Sens and RD53A readout chips were made. Irradiation with protons and neutrons, as well was an extensive test beam campaign at DESY were carried out. 
This paper presents the investigation of various assemblies mainly with ROC4Sens readout chips.
It demonstrates that 
multiple designs fulfill the requirements
in terms of breakdown voltage, leakage current and efficiency. The single 
point resolution for \num{50x50}~\si{\square\micro\meter} pixels 
is measured as \SI{4.0}{\micro\meter} for non-irradiated samples, 
and \SI{6.3}{\micro\meter} after irradiation to $\Phi_{\text{eq}} = \SI{7.2e15}{\per\square\centi\meter}$.

\end{abstract}

\begin{keyword}
Pixel \sep Silicon \sep Sensors \sep CMS \sep HL-LHC \sep Radiation hardness.
\end{keyword}

\maketitle
\section{Introduction}
\label{sec::intro}
To increase the potential for discoveries at the 
Large Hadron Collider (LHC) after Run~3,
a significant luminosity increase of the accelerator is 
targeted~\cite{bruning2019}. CERN therefore plans to upgrade the 
machine to the high-luminosity configuration (HL-LHC) during the Long 
Shutdown~3 (LS3), scheduled for the years 2026-28, 
with the goal of achieving a peak luminosity of
\SI{5.0e34}{\per\square\centi\meter\per\second} nominal, or even 
\SI{7.5e34}{\per\square\centi\meter\per\second} in the ultimate performance scenario
assumed in the following. 
The machine is expected to run at a center-of-mass energy of \SI{14}{\tera\eV} 
with a bunch-crossing separation of \SI{25}{\nano\second} and a 
maximum average of 
200~collisions (pileup) per bunch crossing.
For an expected 10~year operation of the HL-LHC, the CMS experiment
aims to collect
an integrated luminosity of \SI{4000}{\per\femto\barn}. 
To maintain or even improve the performance of
CMS in this harsh environment, the detector will undergo 
several upgrades during the next years. 
In particular, the entire Inner Tracker (IT), which is based on silicon pixel modules, 
will be replaced~\cite{CMSTDR:2017}.

The IT will consist of four barrel layers (TBPX) and twelve forward disks (TFPX and TEPX),
which themselves consist of up to 5 rings, 
at each end of the barrel to extend tracking to a pseudorapidity |$\eta$| = 4.
The innermost barrel layer has a radius of \SI{3.0}{\cm},
while for the other layers the radii are \SIlist{6.1; 10.4; 14.6}{\cm}.
The layers and disks are composed of modular detector units, consisting
of silicon pixel sensors bump bonded to readout chips.
In order to simplify detector
construction and integration and to minimize the number of required spares, 
only two types of detector modules are foreseen, 
namely modules with $1\times 2$ and modules with $2\times 2$~readout chips.

In the innermost pixel layer, a fluence of
particles corresponding to a non-ionising energy loss (NIEL) of a
\SI{1}{\mega\electronvolt} neutron
equivalent fluence of $\Phi_{\text{eq}} = $ \SI{3.5e16}{\per\centi\meter\squared} and a total ionizing dose (TID) of \SI{19}{\mega\gray} will be reached after ten years 
of operation.
To cope with these radiation levels, a 
readout chip using the TSMC 65~nm CMOS technology~\cite{TSMC} is under
development within the RD53 Collaboration~\cite{RD53Aspces:2015}.
The chip will have a non-staggered bump bond pattern
with \SI{50}{\micro\meter} pitch,
which allows a reduction of the pixel area by a factor of six compared
to the current detector, thus improving the spatial resolution and reducing 
the cluster merging, e.g.\ in boosted jets or due to pileup events.
For the studies presented in this paper, an R\&D readout chip is used, the ROC4Sens~\cite{Wiederkehr:2018mox}, 
which is introduced in Sec.~\ref{sec::sensor_des::ROC4Sens}.

Radiation induced bulk damage leads to an increase of leakage current, 
changes of
the electric field and a signal reduction due to charge carrier 
trapping~\cite{klanner2020,Moll:2018bf}. 
Planar silicon pixel sensors are the baseline choice for the entire pixel detector except for 
the innermost barrel layer, where 3D sensors are chosen
due to their higher radiation tolerance and lower power dissipation.
The maximum fluence for planar sensors will be reached in ring~1 of TFPX. 
For the full lifetime of the IT, with \SI{4000}{\per\femto\barn} delivered, the fluence in this ring is expected to reach \SI{2.3e16}{\per\centi\meter\squared}, while in ring~2 of TFPX and 
barrel layer~2 fluences of \SI{1.1e16}{\per\centi\meter\squared} and \SI{9.4e15}{\per\centi\meter\squared} are expected, respectively. The IT is constructed such that ring~1 in 
TFPX could be exchanged after half of the lifetime, which would result in a maximum fluence of 
about \SI{1.2e16}{\per\centi\meter\squared}. At the time of writing it has not yet been decided 
whether TFPX ring 1 will be exchanged. It should also be noted that the fluence in the endcaps depends 
strongly on the radial distance from the beam line. The above quoted numbers refer to the maximum 
fluence, received at the inner module edge, while the mean fluence over the module is much lower, 
about \SI{1.3e16}{\per\centi\meter\squared} over the full detector lifetime. The CMS readout chip 
has been tested up to a total ionizing dose of 10~MGy. Tests at the dose level of 15~MGy, 
expected for the detector region equipped with planar sensors for the full detector lifetime, are 
planned for 2023.
This paper focuses on the characterization of planar silicon pixel sensors for fluences up to the 
maximum expected in a scenario with exchange of TFPX ring 1, namely $\Phi_{\text{eq}} = \SI{1.4e16}{\per\square\centi\meter}$. 
For this, pixel sensors with an active thickness of \SI{150}{\micro\meter} are required 
to achieve a hit efficiency of at least 99\%, with a signal to threshold ratio of 3 or more.


The best spatial resolution is achieved when the projected charge is distributed 
over two pixels.
The CMS Inner Tracker operates in a magnetic field of \SI{3.8}{\tesla},
which results in a strong Lorentz deflection in the direction orthogonal
to the magnetic field $\Vec B$ and the electrical field $\Vec E$,
distributing the signal over two or more pixels in the barrel layers.
For example, for a sensor thickness of \SI{150}{\micro\meter} and a Lorentz angle of \ang{25} this 
deflection amounts to \SI{70}{\micro\meter}.
This means that for pixels with a pitch of \SI{25}{\micro\meter} the 
Lorentz angle has to be reduced by decreasing the mobility, which in turn 
requires a higher electrical field. For the configuration of thickness and pitch mentioned above, a 
straightforward estimate using the relationship 
between field-dependent mobility and Lorentz drift 
yields a bias voltage of about \SI{300}{\volt} in the case of $n^+$-$p$ sensors. 

Overall, the sensor concept must allow for: 
a) operation at high bias voltage without electrical breakdown before irradiation, 
b) operation at up to 800~V to achieve the required hit efficiency after irradation, and 
c) operation without sparking between chip and sensor.

This paper presents the R\&D program for planar silicon pixel sensors 
produced by Hamamatsu Photonics K.K.\ (HPK)~\cite{HPK} with the aim of obtaining 
sensors that meet the criteria for the CMS Inner Tracker as given
in Table~\ref{tab:CMSSpecs}.
\begin{table}[htb]
  \centering
  \caption{Selected requirements for planar pixel sensors~\cite{Steinbruck:2020db}. The full depletion voltage and
  hit efficiency are denoted by $V_{\text{depl}}$ and hit $\epsilon$, respectively.}
   \begin{tabular}{@{}lll@{}}
   \toprule
   Parameter & Value & Measured at \\
   \midrule
    Polarity &  $n^+$-$p$ &\\
    Active thickness & \SI{150}{\micro\meter} & \\
    Breakdown voltage & $\ge$ 300 V & non-irradiated \\
    Breakdown voltage & $\ge$ 800 V & > \SI{5e15}{\per\square\centi\meter} \\
    Leakage current \\ at $V_{\text{depl}}$ + 50 V & 
    $\le$ \SI{0.75}{\micro\ampere \per\square\centi\meter} & \\
    Leakage current \\ at 600 V & 
    $\le$ \SI{45}{\micro\ampere \per\square\centi\meter} & > \SI{5e15}{\per\square\centi\meter} \\ 
     Hit $\epsilon$, before irradiation & $\ge$ 99\% & $V_{\text{depl}}$ + 50 V \\  
     Hit $\epsilon, <$ \SI{1e16}{\per\square\centi\meter} & $\ge$ 99\% & 
     $\le$ \SI{800}{\volt}, \SI{-20}{\celsius} \\  
     Hit $\epsilon, >$ \SI{1e16}{\per\square\centi\meter}  & $\ge$ 98\% &
     $\le$ \SI{800}{\volt}, \SI{-20}{\celsius} \\            
   \bottomrule
  \end{tabular}
  \label{tab:CMSSpecs}
\end{table}

The paper is structured as follows. In Section~\ref{sec::sensor_des}
a detailed description of the pixel sensor layout is given. The sample 
preparation including irradiations is described in Section~\ref{sec::sample}.
The beam test setup and data analysis are presented in Sections 
\ref{sec::beamttest} and \ref{sec::data}. Finally, the results and conclusions
are reported in Sections \ref{sec::res} and \ref{sec::concl}.

\section{Sensor description}
  \label{sec::sensor_des}
A brief and preliminary outline of the first sensor production 
of planar pixel sensors by HPK for this project
can be found in Ref.~\cite{Schwandt:2019lq}. 
In the following, a more comprehensive overview is given.

\subsection{Technological choices}
  \label{sec::senor_des::techno}
The goal of this production was mainly to evaluate different 
silicon substrates and to optimise the pixel layout. For this purpose, 
different types of $n^+$-$p$ sensors were produced on a total
of 35 high-resistivity \SI{150}{\milli\meter} (6'') $p$-type float zone
wafers with crystal orientation <100>. 
The decision for $n^+$-$p$ sensors instead of $n^+$-$n$ 
used in the current CMS pixel detector is not
based on higher radiation hardness 
(after type-inversion the performance of both types is similar),
but on the fact that $n^+$-$p$ sensor production requires
only a single-sided lithography and therefore is potentially cheaper and 
offered by more vendors. An inherent disadvantage of this approach is 
the risk for sparks to form between the sensor edges and the readout chip at 
high voltages (Section \ref{sec::sample::flip}). To solve this issue, 
additional processing steps during bump bonding or module production are 
needed, which partially reduces the advantages of the $n^+$-$p$ approach.

The active thickness of the wafers is chosen to be \SI{150}{\micro\meter}.
For sensors with this thickness, a minimum ionising particle creates about 
\num{11000} electron-hole pairs (most probable value)~\cite{Bichsel1988}.
A reduction by 60\% is expected after the fluence collected in 10 years of operation,
leading to an expected charge of 4400~electrons. 
As the final chip is designed to 
work with an in-time threshold of around \num{1200}~electrons and with built-in data 
sparsification, the module would still have a signal/threshold ratio of about 3 for barrel layers 2-4 and for the disks at the end of operation.

To fabricate the pixel sensors three substrate options have been investigated: 
\begin{enumerate}
\item float zone thinned (FTH150),
\item float zone Si-Si direct bonded (FDB150), 
\item and float zone deep diffused (FDD150). 
\end{enumerate}


The production of the FTH150 
material starts with the same material
and thickness as HPK's standard thick sensors, which is a \SI{320}{\micro\meter} thick float zone 
with 
an approximately \SI{30}{\micro\meter} thick backside implant. After
most of the frontside processing, the backside is mechanically thinned down to the final thickness. Since the frontside has already been processed, there is a 
limitation on the temperature and annealing time for the backside implant
to avoid deformation of the front junction, so that 
the backside implant is much shallower compared to HPK's standard sensors. 
As a result, the backside of these sensors has a higher sensitivity
to scratches, which can lead to a high leakage current in case the 
depletion region touches the backside. 
The effect of such high leakage currents on the module production of large sensors must be evaluated. 

The FDB150 
material is obtained by bonding together 
two wafers: a high resistivity float zone wafer and
a low resistivity handle wafer, which is usually manufactured with the Czochralski method.
The float zone wafer is thinned down to an active thickness of
\SI{150}{\micro\meter}.
After processing the handle wafer is thinned down to \SI{50}{\micro\meter}, resulting in a total thickness of \SI{200}{\micro\meter}.
Compared to the FTH150 wafers, 
the FDB150 wafers are more expensive to produce but less sensitive 
to scratches and handling, which should lead to a higher module yield.

The processing of the FDD150 
material is similar to
the processing of standard float zone material, but with a much deeper backside implant. 
Due to this deeper implant, a more gradual transition from the low-resistivity
to the high-resistivity bulk is achieved compared to the direct-bonded or thinned material~\cite{adam2020a}. 
The diffusion parameters are chosen such that an active layer of
\SI{150}{\micro\meter} is reached and then the wafer is thinned 
down to \SI{200}{\micro\meter}. It is known 
that deep diffusion can introduce material defects~\cite{Junkes:2011jpa}
and possibly dislocations during processing, which can lead to radial as 
well as axial non-uniform doping distributions. 

On the $n^+$-side of the sensor, which is the structured 
electrode side, an inter-pixel isolation is required to isolate neighbouring pixels.
For this production, both $p$-stop and $p$-spray isolation were considered as options.  
For the $p$-spray isolation, a maskless process was chosen,
which, in contrast to the moderated $p$-spray technique used for the current CMS barrel 
pixel sensors~\cite{allkofer2008}, does not require an additional mask. Since HPK 
prefers the $p$-stop technique for reasons of production reliability, 
only a few wafers were produced with the $p$-spray option.

The bulk resistivity was specified to be 3-5~k$\Omega\cdot$cm. All wafers were
processed with a metal grid on the backside to allow light injection.
A summary of the wafer specifications is given in Table~\ref{tab:Waferpar}.
\begin{table}[htb]
  \centering
  \caption{Wafer specifications.}
   \begin{tabular}{@{}ll@{}}
   \toprule
   Parameter & Value \\
   \midrule
    Silicon wafer diameter & \SI{150}{\milli\meter} (6'')\\
    Wafer type & $p$-type, float zone (FZ)  \\
    Crystal orientation & <100> \\
    Active thickness & \SI{150}{\micro\meter} \\
    Total thickness & \SI{200}{\micro\meter} (FDB/FDD), \SI{150}{\micro\meter} (FTH)\\    
    Resistivity & 3 -- 5 k$\Omega\cdot$cm \\
    Oxygen concentration\ & $0.1$ -- $6.5\times 10^{17}$
    \si{\per\cubic\centi\meter} \\
    Number of FTH wafers & 10 ($p$-stop) \\
    Number of FDB wafers & 10 ($p$-stop) + 10 ($p$-spray) \\    
    Number of FDD wafers &  5 ($p$-stop)  \\    
  \bottomrule
  \end{tabular}
  \label{tab:Waferpar}
\end{table}

\subsection{Mask layouts}
 \label{sec::sensor_des::mask}
\begin{figure}[htb]
   \centering
   \includegraphics[width=0.99\linewidth]{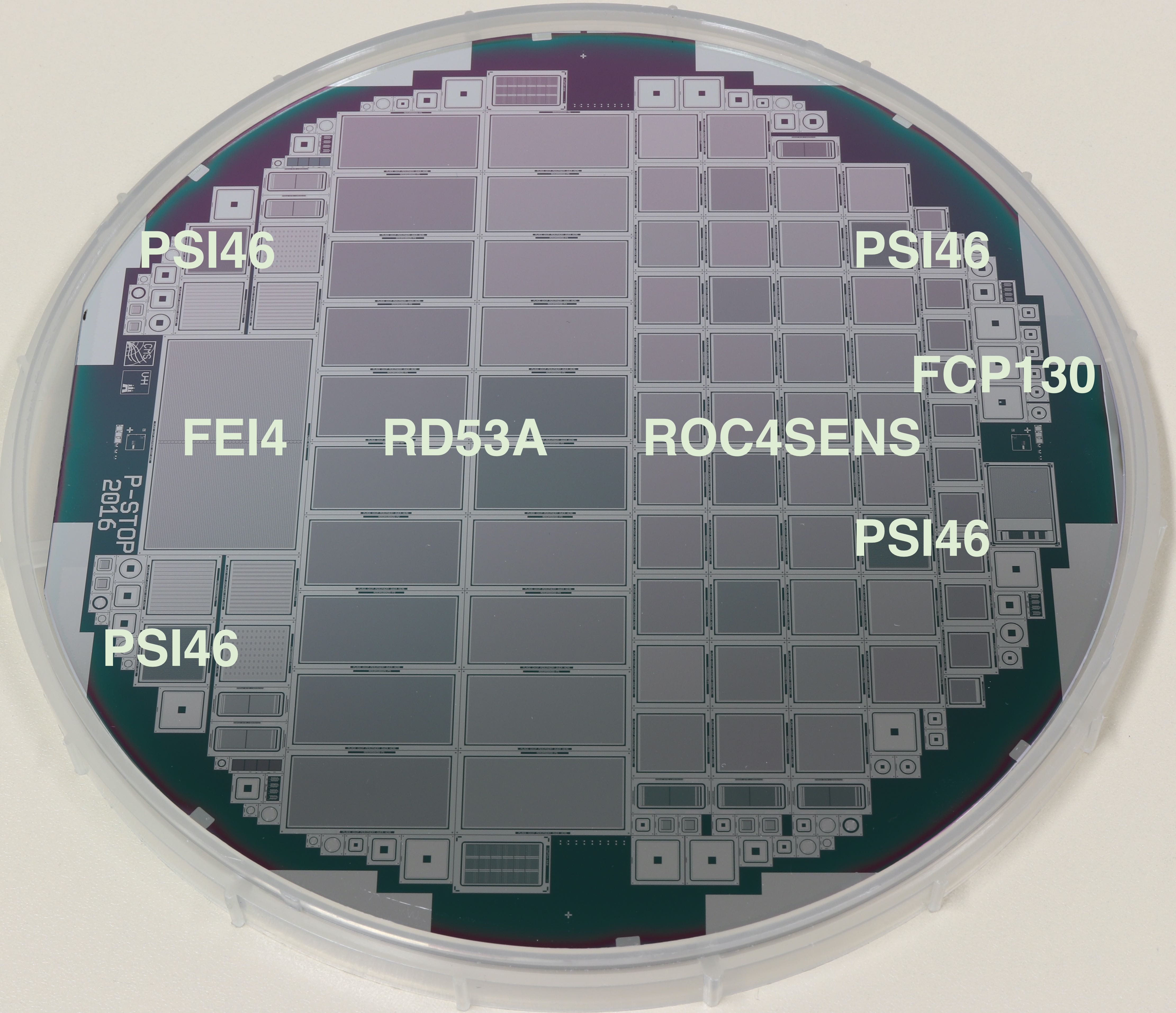}
   \caption{Layout of a \SI{150}{\milli\meter} (6'') HPK sensor wafer with $p$-stop isolation. A wafer includes 20 sensors for the RD53A chip and 39 sensors for the ROC4Sens chip.}
  \label{fig:HPK_Wafer}
\end{figure}

Two different mask sets were produced, one for the wafers with $p$-stop
isolation and one for the wafers with $p$-spray isolation. Each mask set 
contains designs of pixel sensors compatible with different readout chips (bond patterns)
and a variety of test structures, such as diodes of different sizes and shapes, 
MOS-capacitors, MOSFETs and gate-controlled diodes. A picture of a fully processed
$p$-stop wafer is shown in Fig.~\ref{fig:HPK_Wafer}.

As neither the ROC4Sens nor the RD53A chip, both with 
\SI{50}{\micro\meter} pitch (see below),  were available at the time of wafer 
design, sensors compatible with the PSI46 chip~\cite{kastli2006}, which has a bump bond 
pattern of \num{150x100}~\si{\square\micro\meter}, and sensors compatible with 
the FE-I4 chip~\cite{garcia-sciveres2011}, whose bump bond pattern is
\num{250x50}~\si{\square\micro\meter}, were processed as fallback options. 
The sensors designed for the FE-I4 chip were implemented as one double sensor (compatible with two chips)  
in the $p$-stop mask, and as two single sensors in the $p$-spray mask.
Sensors compatible with the PSI46 chip were designed with the default readout 
pattern of \num{150x100}~\si{\square\micro\meter}, but also with
a metal routing structure which allows reading out
\num{100x25}~\si{\square\micro\meter}
and \num{50x50}~\si{\square\micro\meter} subcells.
Since these structures were not bump bonded to readout chips, 
these designs will not be discussed further in the following.

To achieve a high yield during module production, only sensors that fulfil (before irradiation) the specifications given in Table~\ref{tab:CMSSpecs} should be used. In order to obtain meaningful results from a current-voltage ($I$--$V$) measurement of a pixelated sensor 
on the wafer before bump bonding, a bias structure is required to keep all pixel cells on the same potential. After testing, the bias structure
is in general not needed anymore and one aim of this production is to find a 
bias structure that has a minimal impact on the charge collection
and is compatible with high voltage operation after irradiation. For this purpose, sensors 
with common punch-through (PT) structures, polysilicon resistors, open $p$-stop structures, 
and without biasing scheme have been designed. The implementation of the polysilicon 
resistors requires two extra mask layers.
The designs are similar to the sensors described in Ref.~\cite{unno2016} using bias rails
made of polysilicon material.

\subsubsection{Sensor designs for the ROC4Sens readout chip}
  \label{sec::sensor_des::ROC4Sens}
The ROC4Sens is an R\&D chip developed at PSI~\cite{Wiederkehr:2018mox} 
with a staggered bump bond 
pattern of \num{50x50}~\si{\square\micro\meter} and 155 $\times$ 160 channels.
The staggered bump bond pattern is ideal for sensors with 
\num{100x25}~\si{\square\micro\meter} cell size as no metal routing from the 
implants to the bumps on the sensors is required. In case of the $p$-stop mask,
eight different sensors with a cell size 
of \num{100x25}~\si{\square\micro\meter} and nine different
sensors with a cell 
size of \num{50x50}~\si{\square\micro\meter} were designed. For the 
$p$-spray mask, the number of variants was reduced. Common to all designs
is a circular metallisation with a diameter of \SI{20}{\micro\meter}, 
which includes a passivation opening for the bump bond with a 
diameter of \SI{12}{\micro\meter} and the guard ring structure.
 
\begin{figure*}[htp]
   \centering
    \subfloat[]{
     \includegraphics[width=0.4\linewidth]{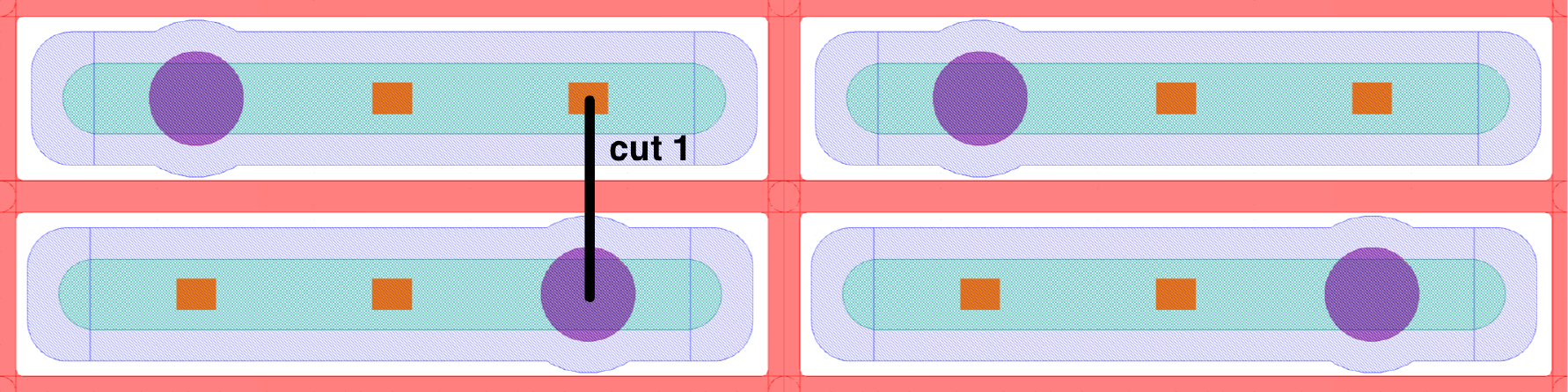}
   }
   \subfloat[]{
     \includegraphics[width=0.4\linewidth]{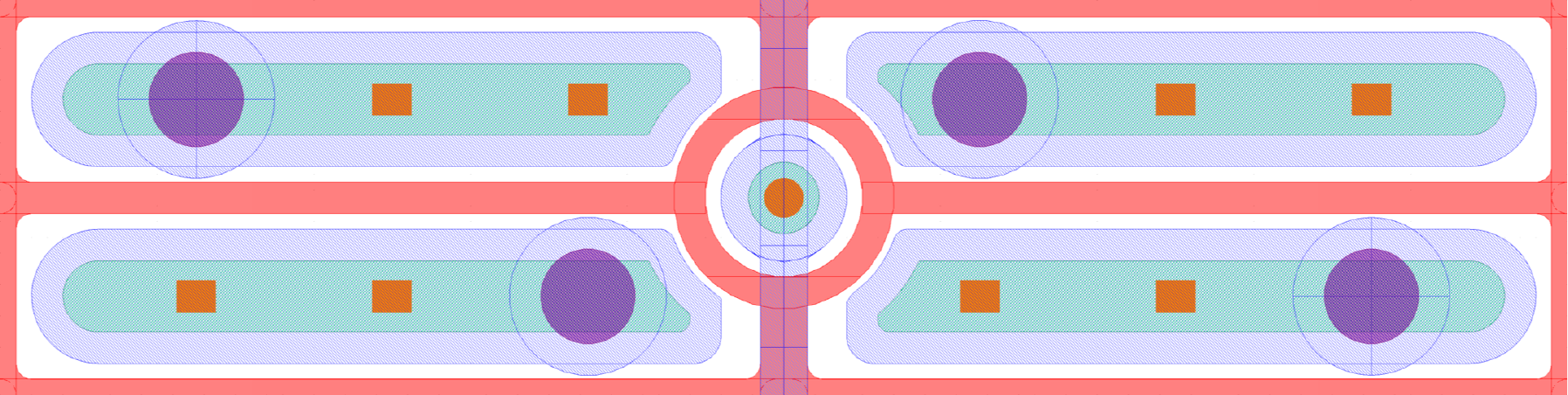}
   }\\
   \subfloat[]{
     \includegraphics[width=0.4\linewidth]{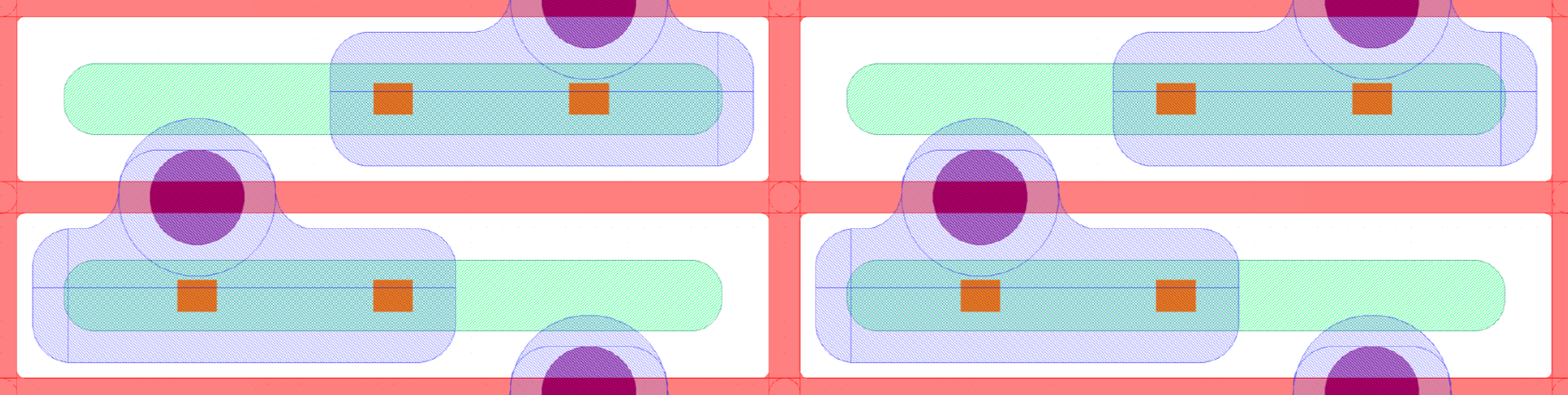}
   } 
    \subfloat[]{
     \includegraphics[width=0.4\linewidth]{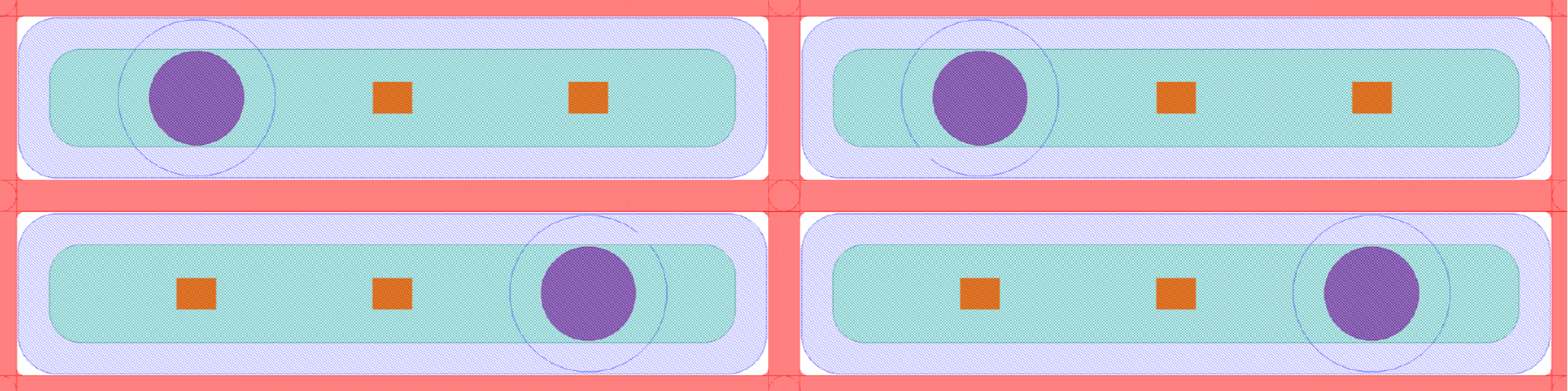}
   }
 
   \subfloat[]{
      \includegraphics[width=0.2\linewidth]{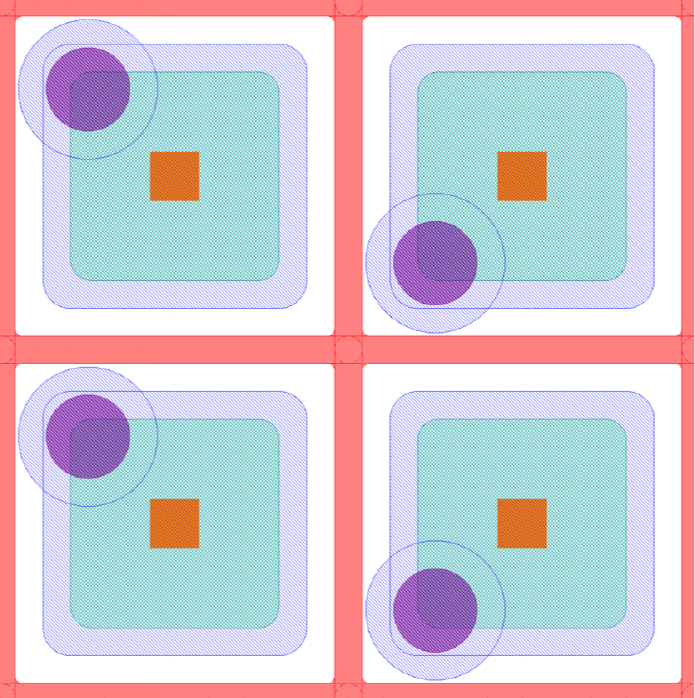}
   }
   \subfloat[]{
      \includegraphics[width=0.2\linewidth]{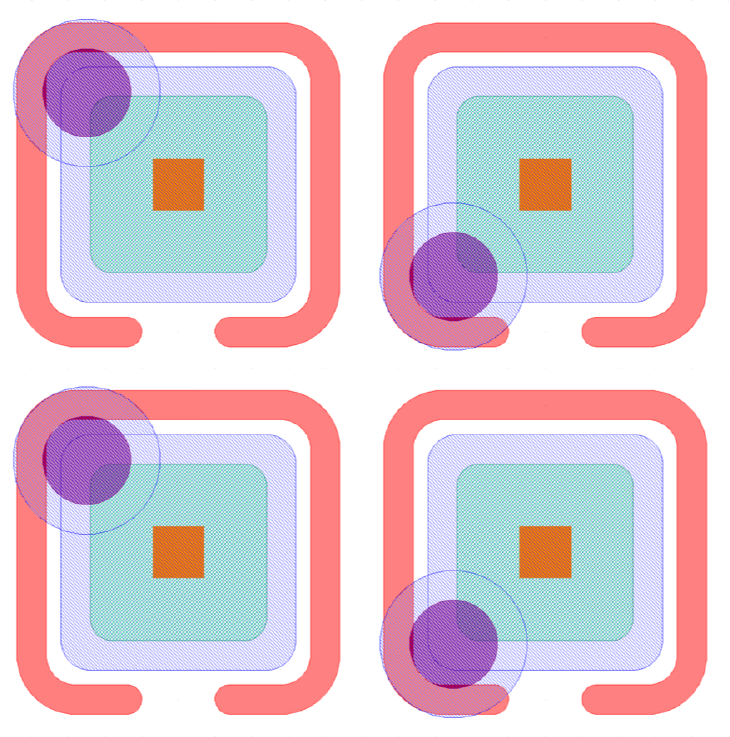}
   }
   \subfloat[]{   
      \includegraphics[width=0.2\linewidth]{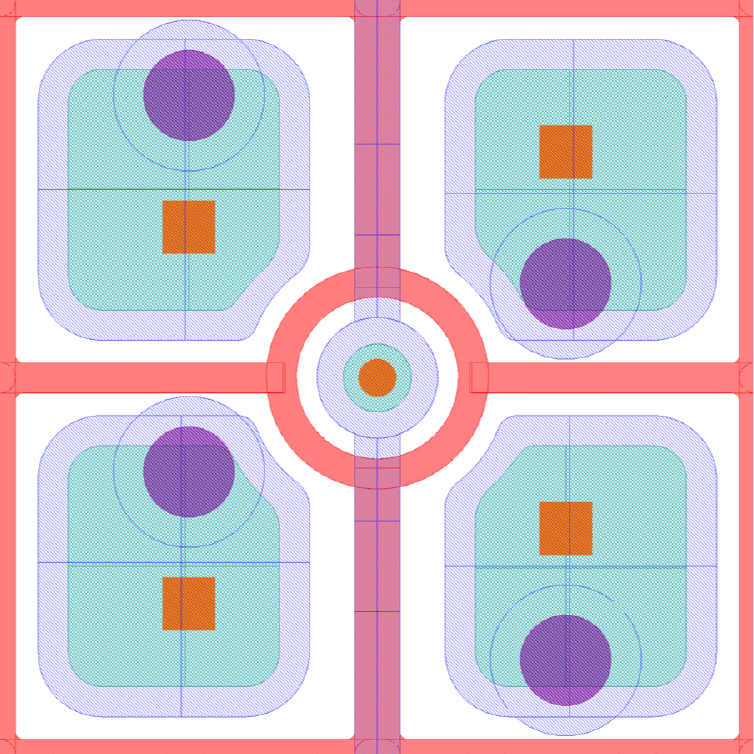}
   } 
    \subfloat[]{   
      \includegraphics[width=0.2\linewidth]{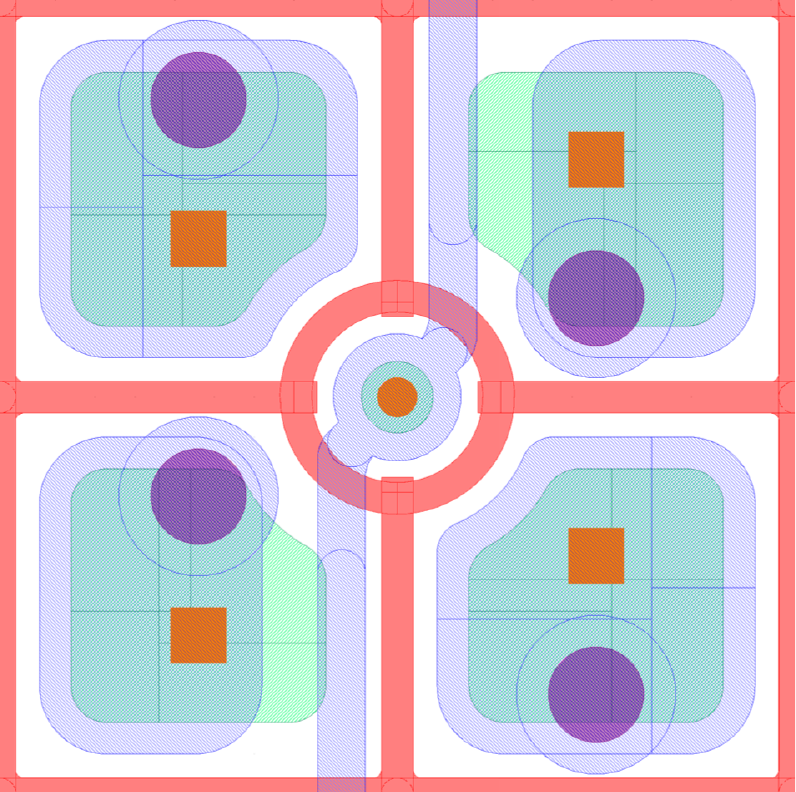}
   }
   \subfloat{
   \includegraphics[width=0.1\linewidth]{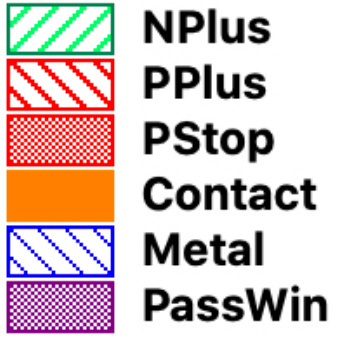}} \\
   \caption{Mask layouts of example designs (\num{100x25}~\si{\square\micro\meter}
   cells in the top two rows and 
   \num{50x50}~\si{\square\micro\meter} cells in the bottom row) for the 
   ROC4Sens chip with $p$-stop isolation:
   a)~Default, no bias scheme (R4S100x25-P1).
   b)~Common punch-through and straight bias rail (R4S100x25-P2).
   c)~Routing test, no bias scheme (R4S100x25-P4).
   d)~Maximum implant, no bias scheme (R4S100x25-P7).   
   e)~No bias scheme (R4S50x50-P1).
   f)~Open $p$-stop (R4S50x50-P2).
   g)~Common punch-through and straight bias rail (R4S50x50-P3).
   h)~Common punch-through and wiggle bias rail (R4S50x50-P4). The color code indicates the various mask layers: $n^+$ implant (NPlus), $p^+$ implant (PPlus), $p$-stop implant (PStop), metal contact via (Contact), metallization (Metal), opening in the passivation (PassWin).}
  \label{fig:R4Sgds}
\end{figure*}

The mask layouts of the most promising pixel cells with 
$p$-stop isolation are shown in Fig.~\ref{fig:R4Sgds}.
These are for the \num{100x25}~\si{\square\micro\meter} cell:
\begin{enumerate}
 \begin{figure}[htbp]
    \includegraphics[width=0.99\linewidth]{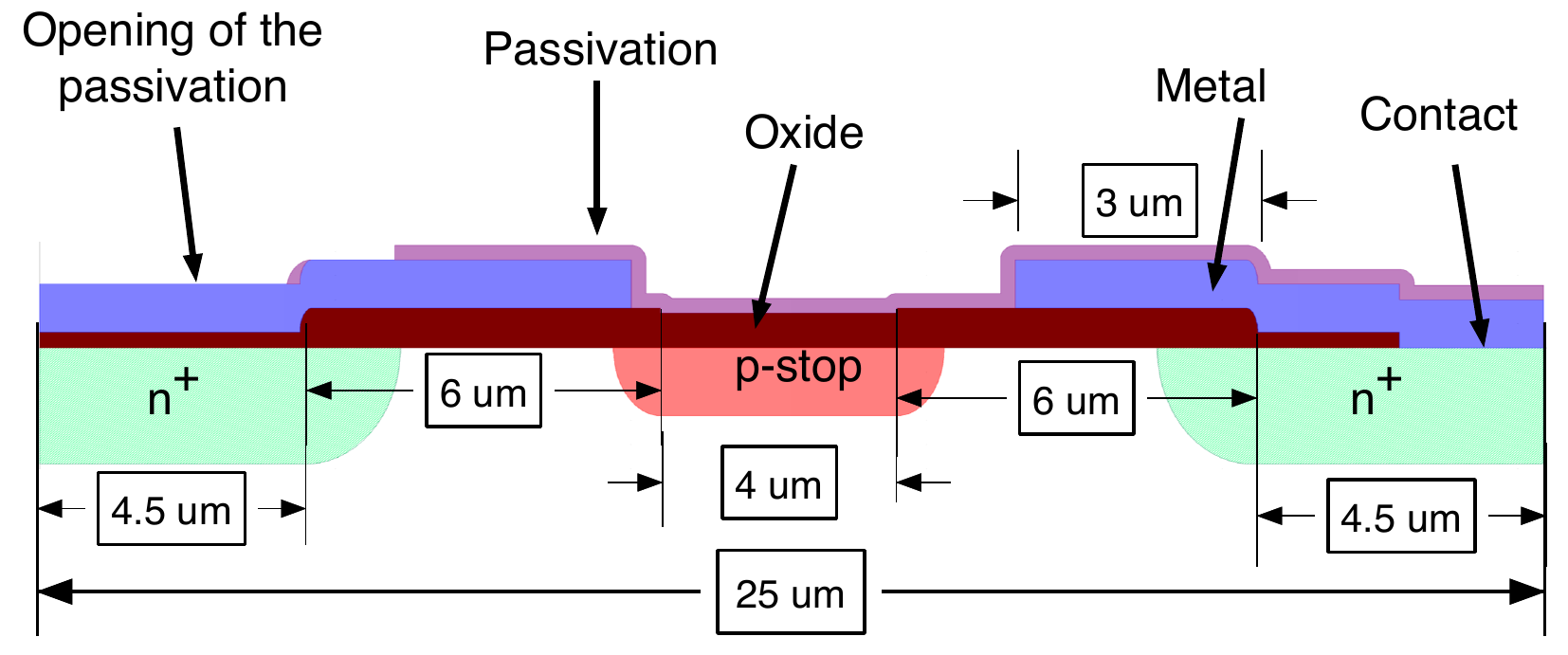}
    \caption{Cross section of the region between two pixels (marked as "cut 1" in Fig.~\ref{fig:R4Sgds}(a)) for a sensor with no bias scheme (R4S100x25-P1).
     Horizontal dimensions are taken from the GDS file, vertical dimensions are only indicative.}
    \label{fig:R4S100x25p-P1cut}
  \end{figure}
  \item[a)] Sensor with no bias scheme (R4S100x25-P1).
  The cross section along the \SI{25}{\micro\meter} direction,
  together with the relevant dimensions of the design, is shown
  in Fig.~\ref{fig:R4S100x25p-P1cut}.
  The width of the $n^+$ implant is \SI{9}{\micro\meter},
  the width of the metal overlap is \SI{3}{\micro\meter} 
  and the $p$-stop implant has a width of \SI{4}{\micro\meter}.
  \item[b)] Sensor with common punch-through for simultaneous biasing of 
  four pixels and a straight bias rail (R4S100x25-P2). The 
  $n^+$ bias dot has a diameter of \SI{10}{\micro\meter},
  which is necessary to form the contact hole within the production tolerance.
  The total diameter, 
  including the surrounding $p$-stop implant, is \SI{30}{\micro\meter}.
  To reduce the losses along the bias rail, the $p$-stop implantation
  underneath is wider than the metallisation of the rail~\cite{unno2016}.
  \item[c)] Sensor with bump bond pad in the middle of two pixels on top of the $p$-stop implant.
  This is used for routing tests (R4S\-100x25-P4).
  \item[d)] Sensor with a wider $n^+$ implant (R4S100x25-P7). The width
  is \SI{12.5}{\micro\meter} and the metal overlap \SI{3}{\micro\meter}, resulting
  in a minimal distance between the metal plates of \SI{5.5}{\micro\meter}.
\end{enumerate}
For the \num{50x50}~\si{\square\micro\meter} cell
the designs are:
\begin{enumerate}
\item[e)] Sensor with no bias scheme (R4S50x50-P1). The
$n^+$ implant is \SI{30}{\micro\meter} wide. 
\item[f)] An open $p$-stop design
with an $n^+$ implant width of \SI{24}{\micro\meter} (R4S50x50-P2).
\item[g)] Sensor with common punch-through for simultaneous biasing of 
four pixels and a straight bias rail (R4S50x50-P3). The $n^+$ implant 
size is \num{28x32}~\si{\square\micro\meter}. The bias dot and the bias rail
are the same as for R4S100x25-P2. 
\item[h)] Sensor with common punch-through and a wiggle bias rail to prevent an overlap 
with the $p$-stop implant. The $n^+$ implant size is 
\num{32x32}~\si{\square\micro\meter}.
\end{enumerate}
In addition, sensors with bias dot, without a gap between the 
$n^+$-dots and the surrounding $p$-stop implant and sensors
with polysilicon resistors have been designed for the ROC4Sens chip.
The non-irradiated sensors with polysilicon functioned electrically,
but exhibited problems in the test beam measurements, 
due to a too low resistance of the resistors.
This manifested itself in a pattern in the hit map with a 
central band of pixels with signals and a cluster charge too small by a factor of two.
Therefore, they are not considered as an
option in the following.

\subsubsection{Sensor designs for the RD53A chip}
\label{sec::sensor_des::RD53A}
The RD53A chip is a prototype chip developed by the RD53 Collaboration 
with a non-staggered bump bond pattern of 
\num{50x50}~\si{\square\micro\meter} 
and 192 $\times$ 400 cells. The non-staggered bump bond pattern makes it 
necessary, in case of the \num{100x25}~\si{\square\micro\meter} pixel size,
to implement a metal routing connecting the $n^+$ implant to the bump.
Such routing on the sensor may result in additional cross talk between adjacent pixels. This issue needs to be further investigated with the RD53A readout chip. 

\begin{figure*}[htbp]
   \centering
   \subfloat[]{
     \includegraphics[width=0.4\linewidth]{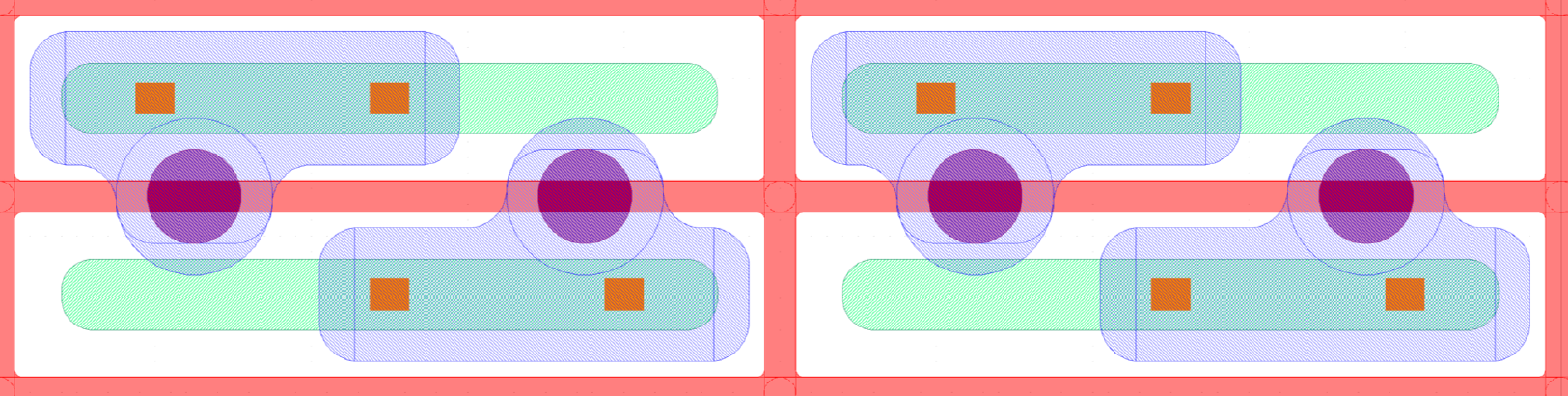}
   }
   \subfloat[]{
     \includegraphics[width=0.4\linewidth]{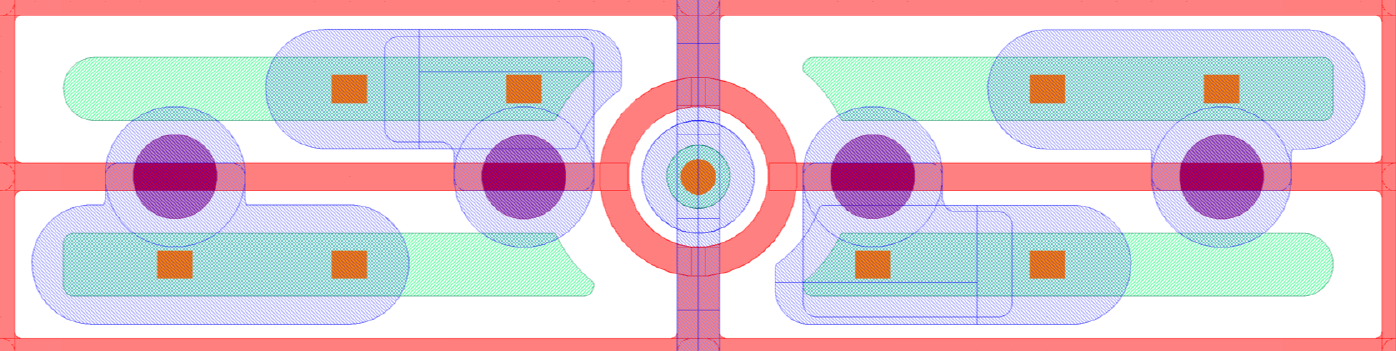}
   } \\
   
   \subfloat[]{
      \includegraphics[width=0.2\linewidth]{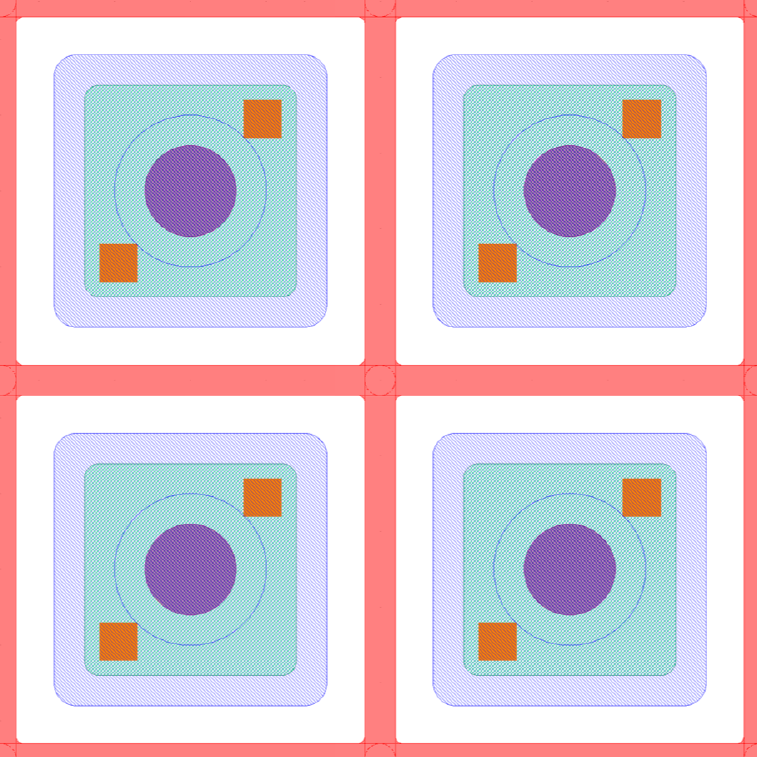}
   }
   \subfloat[]{
      \includegraphics[width=0.2\linewidth]{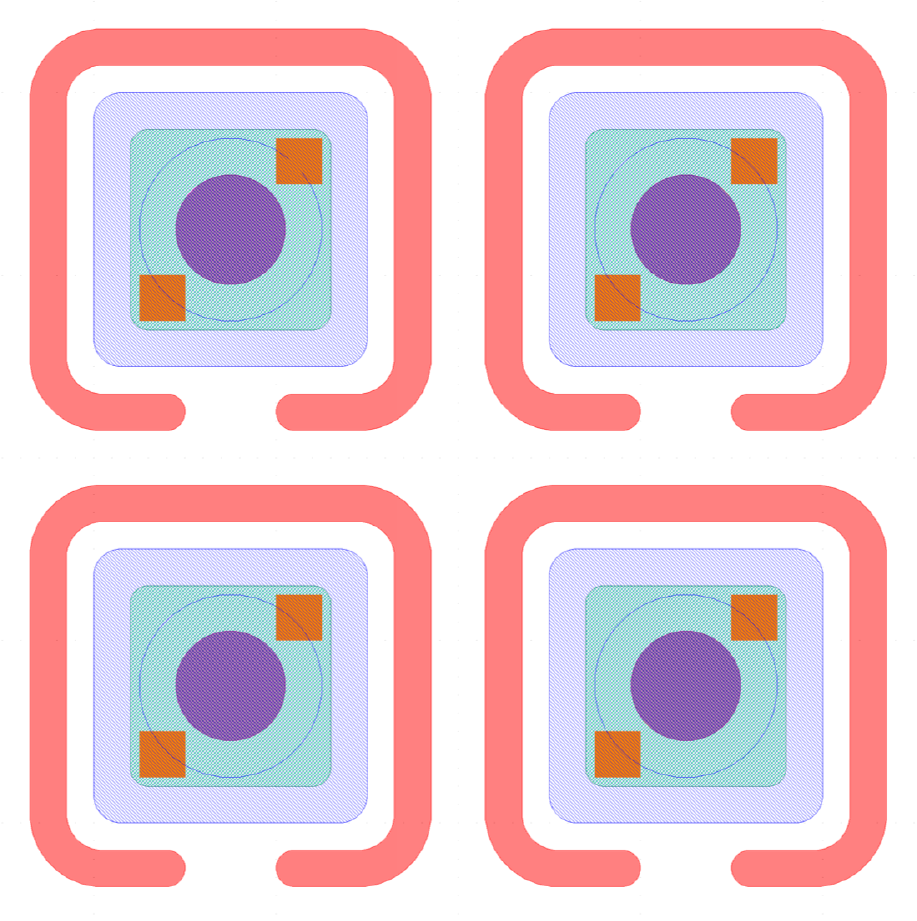}
   }
   \subfloat[]{   
      \includegraphics[width=0.2\linewidth]{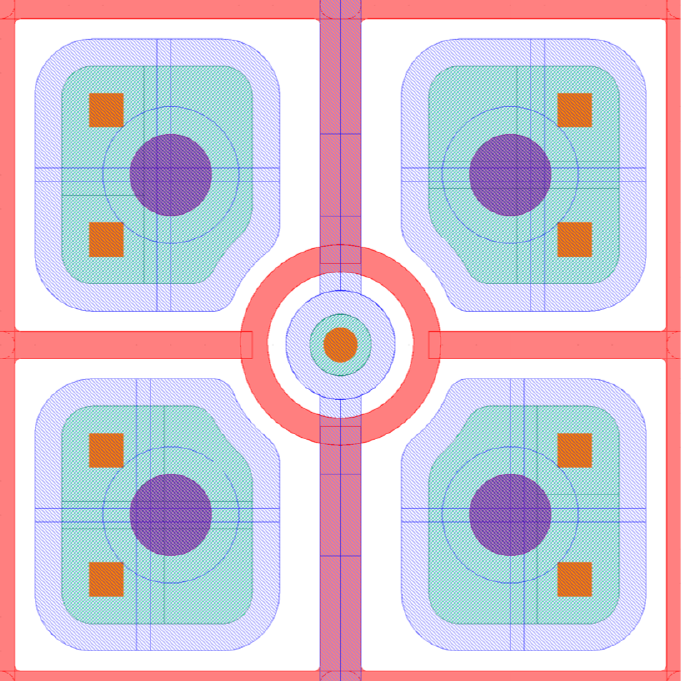}
   } 
    \subfloat[]{   
      \includegraphics[width=0.2\linewidth]{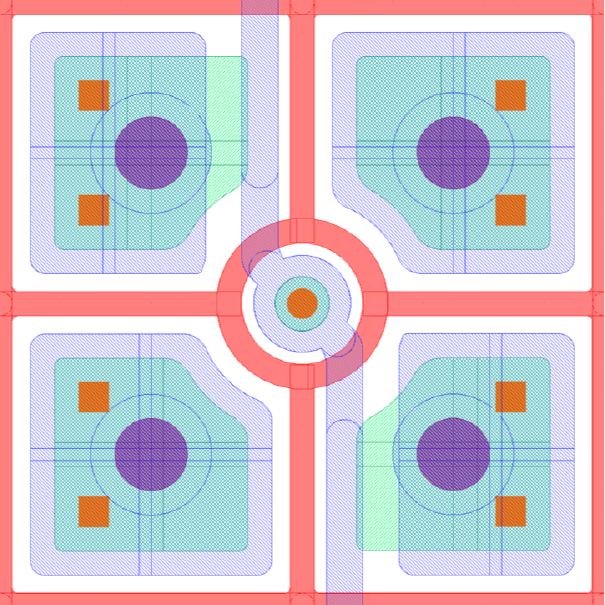}
   }
   \subfloat{
   \includegraphics[width=0.1\linewidth]{figures/gdslayers.pdf}} \\    
   \caption{Mask layouts of example designs (\num{100x25}~\si{\square\micro\meter} cells in the top row and \num{50x50}~\si{\square\micro\meter} cells
   in the bottom row) for the RD53A chip with $p$-stop isolation:
   a)~Default, no bias scheme (RD53A100x25-P1).
   b)~Common punch-through and straight bias rail (RD53A100x25-P2).   
   c)~Default, no bias scheme (RD53A 50x50-P1).
   d)~Open $p$-stop (RD53A50x50-P2).
   e)~Common punch-through and straight bias rail (RD53A50x50-P3).
   f)~Common punch-through and wiggle bias rail (RD53A50x50-P4). The color code indicates the various mask layers: $n^+$ implant (NPlus), $p^+$ implant (PPlus), $p$-stop implant (PStop), metal contact via (Contact), metallization (Metal), opening in the passivation (PassWin).}
  \label{fig:RD53Agds}
\end{figure*}

Twenty sensors (ten variants) for the RD53A chip are placed on a wafer. 
Of these, eight sensors 
have a \num{100x25}~\si{\square\micro\meter} cell and twelve sensors 
have a \num{50x50}~\si{\square\micro\meter} cell. For the 
$p$-stop mask, the mask layout of the most promising designs are shown in 
Fig.~\ref{fig:RD53Agds}.
The dimensions of the $n^+$ implants, $p$-stop implant and bias dots are the same
as for the design for the readout with the ROC4Sens chip.

\subsubsection{Guard ring}
\label{sec::sensor_des::guardring}
\begin{figure}[htb]
    \center
    \includegraphics[width=0.75\linewidth]{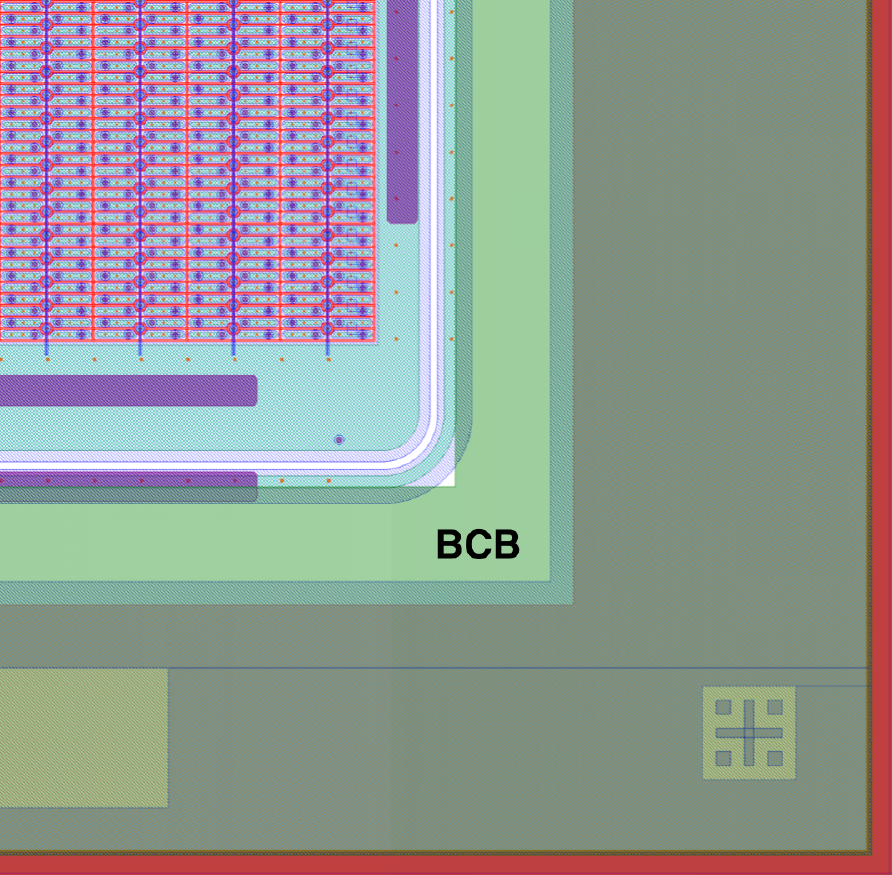}
    \caption{Design of the guard-ring structure of a R4S100x25-P2 sensor 
    including a benzocyclobutene (BCB) mask (green layer) aiming to prevent sparking. The BCB layer is designed as a frame that extends from the outer guard ring to the cut edge.}
    \label{fig:guardring}
\end{figure}

All sensitive sensor areas are surrounded by a guard-ring structure (Fig.~\ref{fig:guardring})  
consisting of an inner or bias ring (in case of a bias structure),
an outer ring and an edge ring. 
The inner and outer rings have openings in the passivation to allow for 
probing with needles.
In addition, there are passivation openings for bumps on each side of the 
bottom of the inner ring that can be connected to the readout chip. 
This allows the inner ring to be either set to ground or left floating.
In the case of a sensor without bias structure, 
grounding the inner ring should result in less noisy edge pixels, 
since the current from the inactive area is drained through this ring. 
The RD53A chip has the possibility of switching between 
both states by a jumper on the readout card, whereas this option is not available with the 
ROC4Sens chip. In this case, the UBM (Under Bump Metallisation) mask defines 
if the inner ring is grounded or left floating.
The following measurements with the ROC4Sens chip
are performed with the inner ring grounded, while for the measurements with 
the RD53A chip the inner ring was left floating. 

\subsection{Electrical measurements \& yield}
\label{sec::sensor_des::electrical}
\begin{figure}[!htb]
   \centering
   \subfloat[]{
     \includegraphics[width=0.9\linewidth]{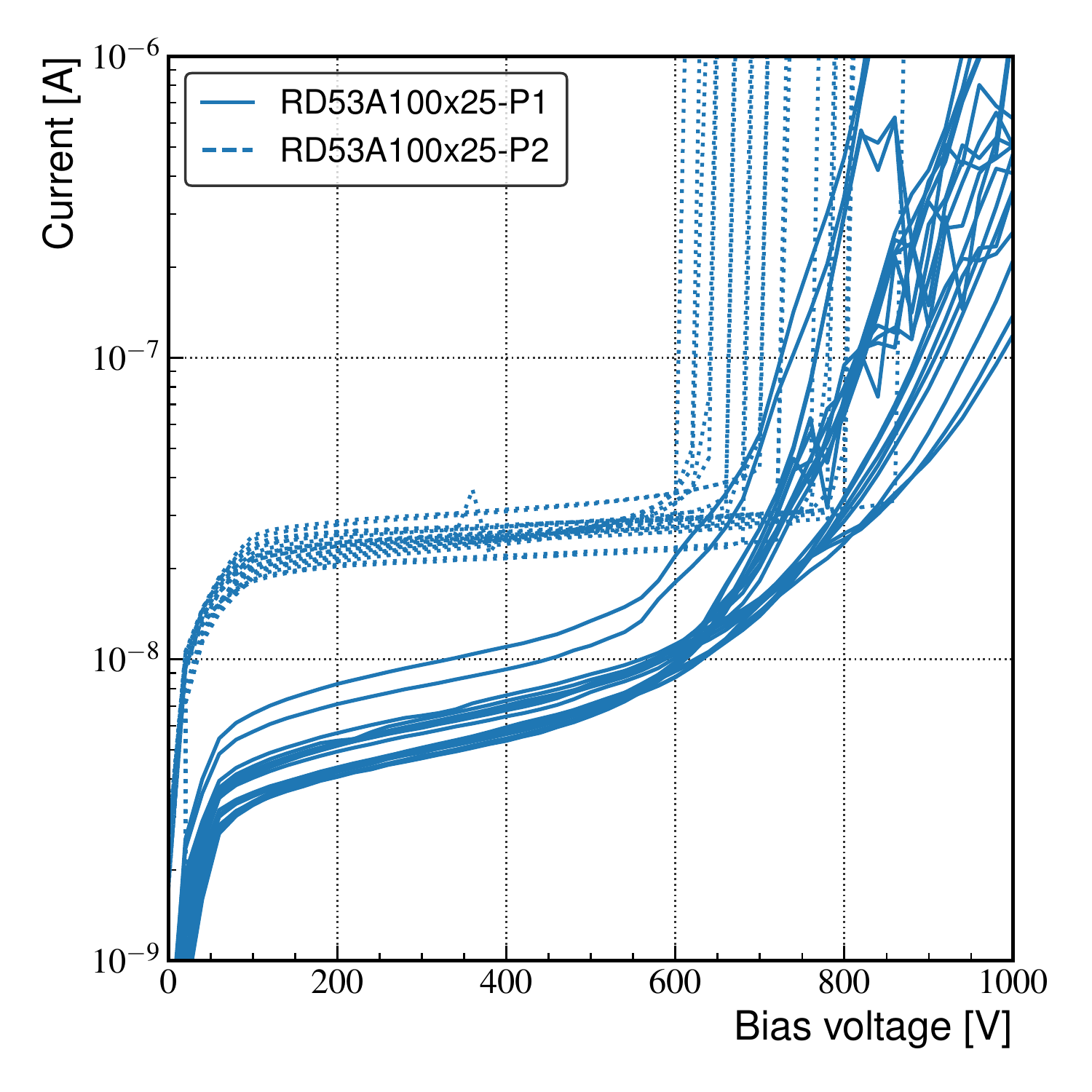}
   }\\
   \subfloat[]{
     \includegraphics[width=0.9\linewidth]{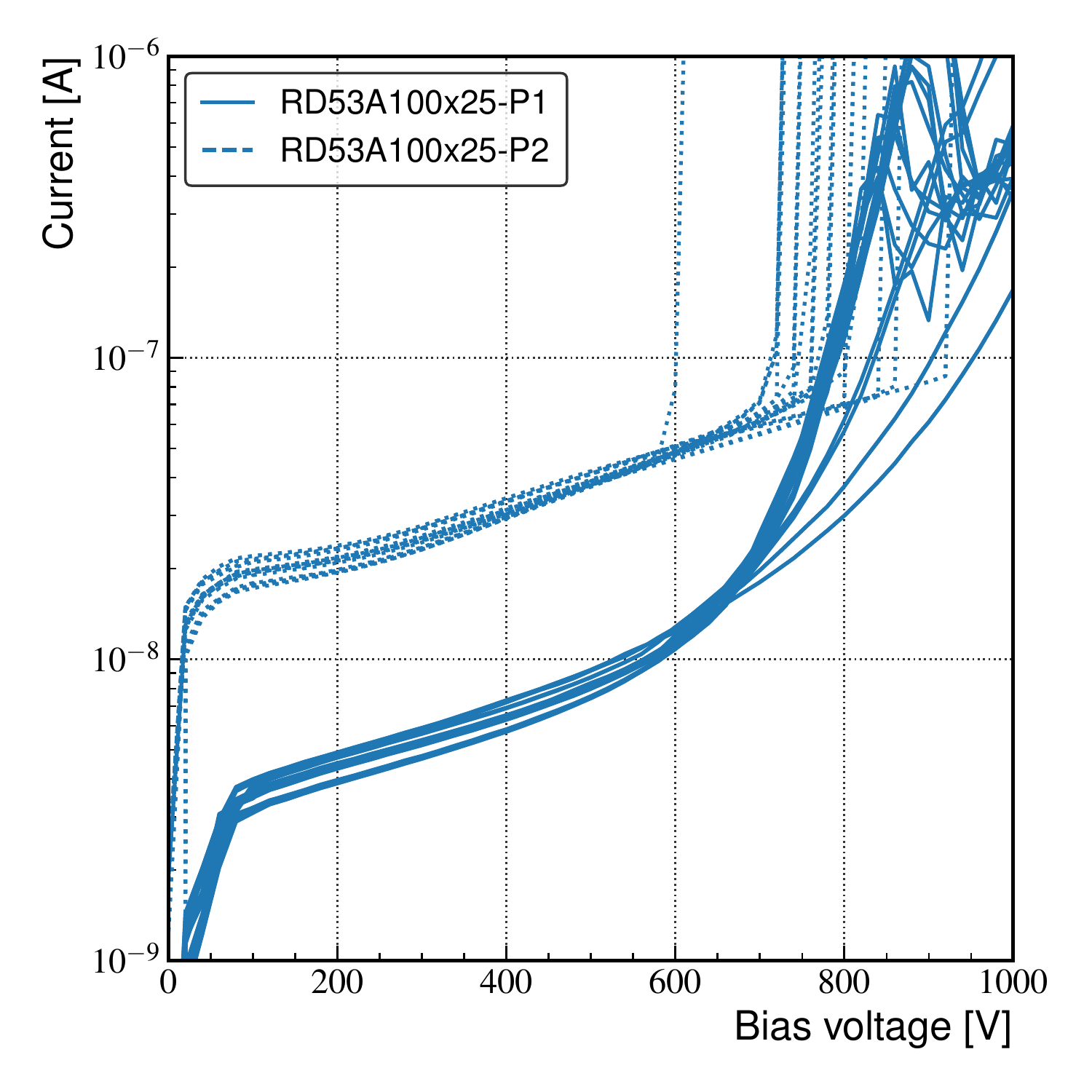}
   } 
    \caption{$I$--$V$ measurements of all RD53A100x25-P1 (no bias scheme, solid lines)
    and RD53A100x25-P2 (with common punch through, dashed lines) sensors
    on (a) FDB150 wafers and (b) FTH150 wafers. Unlike the sensors with bias dot from 
    the FDB150 wafers, the current of the sensors with bias dot from the FTH150 wafers 
    continues to increase even after full depletion.}
   \label{fig:IV_RD53A}
\end{figure}

For an R\&D production with new sensors, it is difficult to define meaningful 
acceptance criteria for the wafer.
Therefore, sensor designs already successfully used during CMS' HPK campaign~\cite{adam2020a}
and pad diodes were used for this production to facilitate the acceptance 
of the wafers. 
Current-voltage measurements were performed by HPK on all sensors 
and diodes on the bias ring and inner guard ring, respectively.
The measurements were done in \SI{20}{\volt} steps up to \SI{1000}{\volt}. 
All delivered wafers met the requirements in terms of 
full depletion voltage, leakage current and breakdown voltage as specified in Table~\ref{tab:CMSSpecs}. 
In general the results indicated a high fraction of acceptable
sensors with high breakdown voltage (> \SI{600}{\volt}) for the 
different sensor designs, but also revealed some problematic combinations 
of sensor design and material. 
For example, on the FDB150 wafers with $p$-stop isolation the sensors of type
R4S100x25-P2 have a yield of only 25\%, while they have a yield of
100\% on the FTH150 and FDD150 wafers. 
It is also observed that the leakage current on the
FDD150 wafers is a factor of 10 larger compared to the FTH150 and FDB150 wafers, 
and it varies significantly across a wafer.
As a consequence, sensors on FDD150 wafers with bias structure cannot 
be distinguished from sensors without bias structure based on $I$--$V$ 
measurements. This is in contrast to the case on FTH150 and FDB150 wafers, whose $I$--$V$ curves are 
shown in Fig.~\ref{fig:IV_RD53A}, and complicates the determination of good FDD150 sensors using the $I$--$V$ measurements.

The reason for the high leakage current of sensors from FDD150 wafers is probably a deep hole trap 
with the designation $H(220K)$, which was found using deep-level transient spectroscopy on 
similarly processed test structures \cite{Junkes:2011jpa} and is known as a possible current generator. In addition, a very high oxygen concentration and a thickness dependence of the defect concentration were found. From this it can be concluded that the defects were formed during the deep-diffusion process.

\begin{figure}[htb]
    \includegraphics[width=0.99\linewidth]{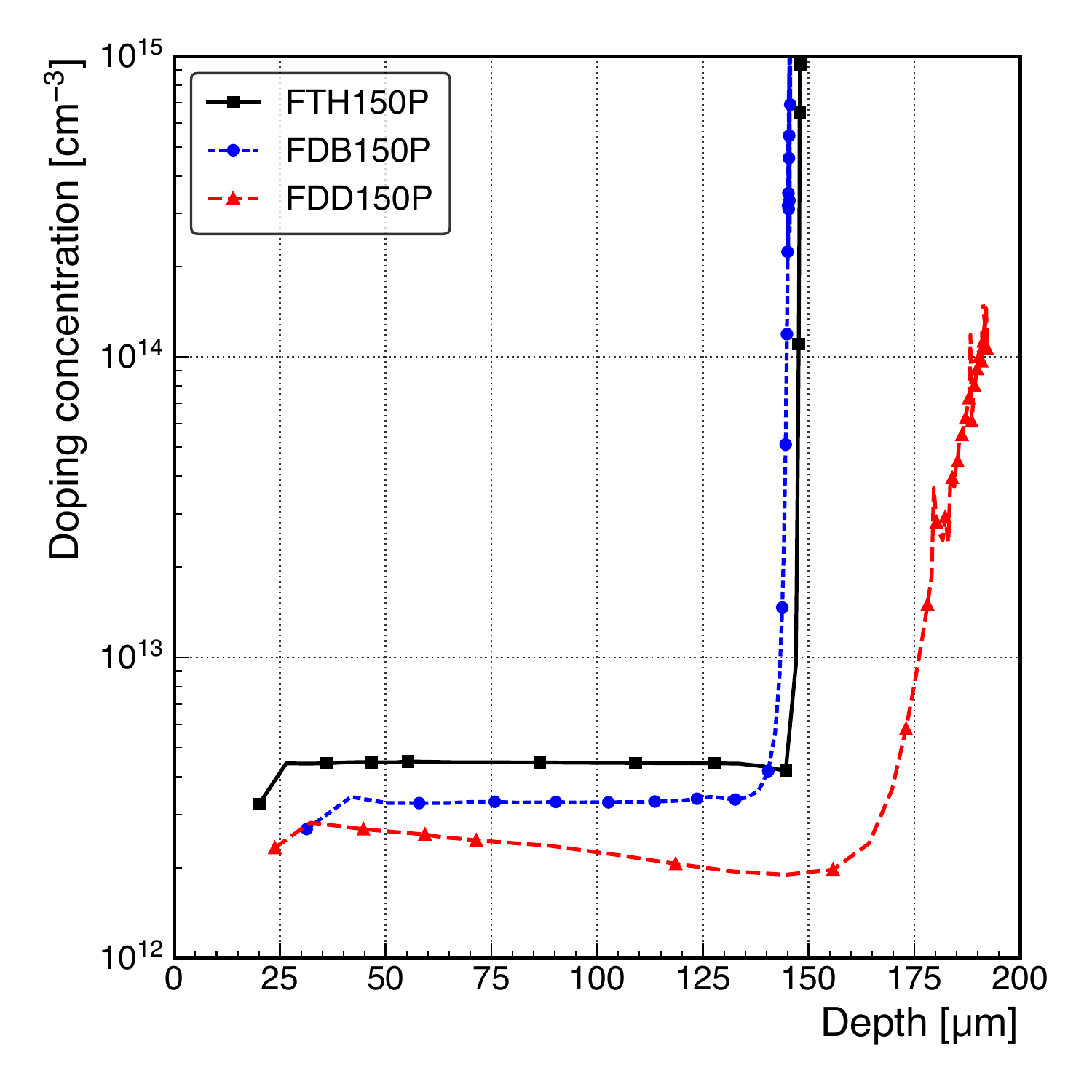}
    \caption{Typical doping profiles for the different $p$-doped substrates extracted from $C$--$V$ measurements on diodes.}
    \label{fig:doping}
\end{figure}

Capacitance-voltage ($C$--$V$) measurements on diodes of different sizes were performed to determine 
the full depletion voltages and doping profiles taking edge effects 
into account~\cite{Fretwurst:2017cy}. The full depletion voltage is 
in the range of 55 to \SI{75}{\volt}, depending on the substrate.
Examples of doping profiles of the different substrates are shown in
Fig.~\ref{fig:doping},
indicating that the active thickness of FTH150 and FDB150 sensors is close to the 
specified \SI{150}{\micro\meter}.
The bulk doping concentration of FTH150 sensors is around 
\SI{4.4e12}{\per\cubic\centi\meter}, while it is 
\SI{3.3e12}{\per\cubic\centi\meter} for FDB150 sensors.
The doping profile of the deep diffused substrate is very inhomogeneous in the
sensitive region of the sensor and the 
active thickness is larger than \SI{175}{\micro\meter}. As a result, 
this material is excluded from further consideration.

\section{Sample preparation}
 \label{sec::sample}
\subsection{Readout chips}
 \label{sec::sample::chip}
Both types of readout chips, the ROC4Sens chip and the RD53A chip, were used
to characterise the HPK sensors.

The ROC4Sens chip is based on the 
PSI46 chip (fabricated in the same IBM \SI{250}{\nano\meter} process) 
and is intended for sensor studies only.
The chip has \num{24800} pixels and a total size of
\num{9.8x7.8}\si{\milli\squared\meter}. The chip is easy to operate
and can be read out with the same Digital Testboard (DTB) as used for the testing
of the CMS Phase-1 pixel readout chips~\cite{CMSPix1_2021} 
after adapting the firmware, adapter and software.
There are no DACs to be set, only two shift registers to be programmed, no discriminator 
threshold adjustment or trimming needed and only a small number of control signals is 
required. The signal processing of each pixel features a pre-amplifier and
a shaper with fast pulse shaping. The collected charge can be stored 
on a sample-and-hold capacitor. When the charge of a hit is being stored, 
the pixel cannot accept further incoming hits. 
As there is no internal signal on the chip or pixel which 
indicates a hit, the storage and readout of a hit has to be 
triggered externally with the trigger signal
distributed to all pixels simultaneously. 
With digitisation of all pixels with \SI{12}{bit} resolution in the DTB this allows for data taking 
without zero suppression at rates of around \SI{150}{\hertz}. 
To save disk space only regions of interest, \num{7x7} pixels centred 
around a seed pixel with a charge above threshold, are stored.  

The RD53A chip~\cite{RD53Aspces:2015} 
is a prototype for the ATLAS and CMS readout chips 
planned for operation at the HL-LHC.
The chip has three analogue front-end flavours. 
Only the linear front-end, which covers 1/3 of the entire 
pixel matrix and which is the front-end selected 
by CMS~\cite{Adam_2021}, is used in this study. 
It provides a self-triggering mode,
which facilitates source scans to be performed, and stores
the charge using the time-over-threshold method with \SI{4}{bit} accuracy. For non-irradiated chips a threshold $\le$~1000~electrons is achieved.
 
\subsection{Flip chip \& spark protection}
  \label{sec::sample::flip}
Under-bump metallisation on the sensor wafer, bump deposition on the 
chip wafer and flip-chip bonding of single-chip ROC4Sens and RD53A 
modules were done at Fraunhofer IZM~\cite{IZM}. 
The technology chosen uses SnAg bumps on the readout chip and 
Ni-Cu pads on the sensor. The chips for the studies of this paper
were \SI{700}{\micro\meter} thick. In case of the
ROC4Sens modules, the bump-bond yield was usually above 99.5\%.

To prevent sparking between sensor and chip at high bias voltage the 
option to use a benzocyclobutene (BCB) frame on the sensor~\cite{Weigell:2011is} has been 
investigated. 
The BCB was deposited as a frame from the cut edge to the bias 
ring on the sensor, as shown in Fig.~\ref{fig:guardring}. 
However, measurements carried out on non-irradiated modules in the laboratory 
showed sparking at a voltage of \SI{490}{\volt}, requiring alternative 
solutions.
For the test beam measurements, it was found that a protection of the
modules with SYLGARD$^{\text{TM}}$~184 Silicone Elastomer~\cite{Sylgard} was sufficient
to safely operate the modules up to \SI{800}{\volt} without sparking.
SYLGARD is not a practical option for the module production of the final detector, but we do not expect its usage to affect the results obtained in this paper. 

\subsection{Irradiations}
  \label{sec::sample::irrad}
At the radial position of the pixel sensors the fluence is dominated by charged hadrons, therefore those should be used in irradiation studies. 
Unfortunately, for higher proton fluences the shaping time in the ROC4Sens chip cannot be configured as needed. 
To achieve fluences above \SI{5.3e15}{\per\square\centi\meter},
the modules were irradiated  with neutrons.
Even though the electrical fields and trapping times are different after proton 
and neutron irradiations~\cite{Kramberger:2019},
it was shown in Ref.~\cite{Affolder:2010} 
that the collected charge in $n^+$-$p$ sensors is similar.

\begin{figure}[htb]
  \centering
  \includegraphics[width=0.75\linewidth]{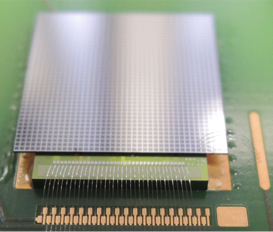}
   \caption{ROC4Sens single chip module mounted on a PCB. The backside metal grid on the
   sensor is to allow laser injection.}
  \label{fig:Roc4Sens_mod}
\end{figure}

Before proton irradiation most of the modules were first glued to a printed circuit board (PCB), wire bonded and tested for basic functionality.
An example module is shown in Fig.~\ref{fig:Roc4Sens_mod}.  
For neutron irradiation, untested bare modules were put into 3D-printed boxes, and irradiated before wire bonding.

A list of all samples used in the following studies is given in Table~\ref{tab:Samples}.

The neutron irradiation was performed in the TRIGA Mark~II reactor in Ljubljana. The \SI{1}{\mega\electronvolt} neutron equivalent fluences $\Phi_{\text{eq}}$ 
were 0.5, 3.6, 7.2 and \SI{14.4e15}{\per\square\centi\meter}, determined using a 
hardness factor of 0.9~\cite{Zontar:1998noa}. 
 
\begin{figure}[htb]
  \includegraphics[width=0.99\linewidth]{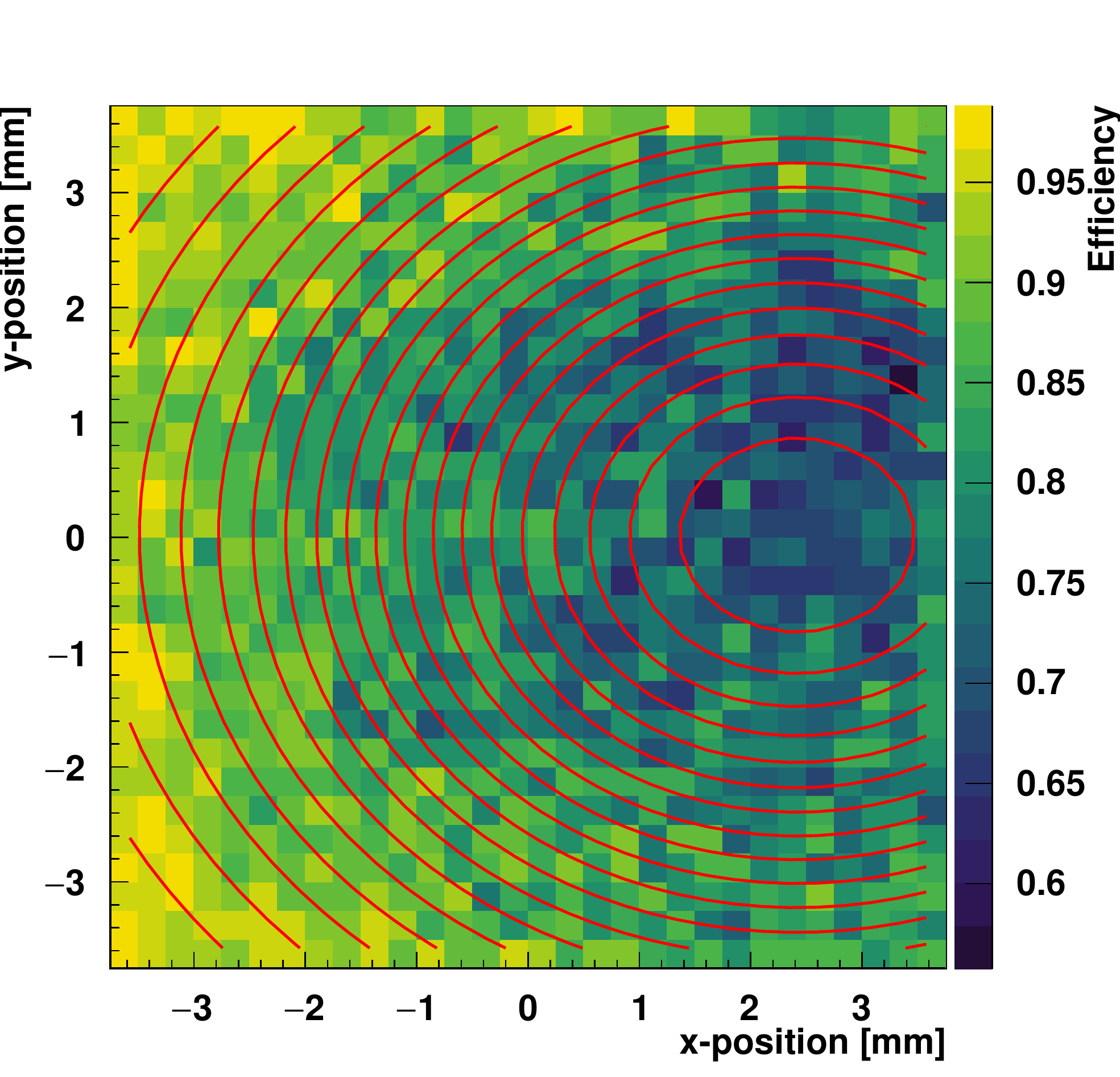}
    \caption{Hit efficiency distribution of a ROC4Sens module measured at \SI{800}{\volt} irradiated with \SI{24}{\giga\electronvolt\per c} protons 
    at CERN IRRAD. 
    Lines of constant efficiency are shown to indicate the
    reconstructed proton fluence profile. It is  
    clearly visible that the module was not centred in the beam.}
      \label{fig::beamspFit}
\end{figure}

The proton irradiation was performed at the PS-IRRAD Proton Facility at CERN 
(PS) with a beam momentum of \SI{24}{\giga\electronvolt\per c} to fluences 
$\Phi_{\text{eq}}$ of 2.0 and \SI{4.0e15}{\per\square\centi\meter} averaged over the sensors.
The hardness factor used in the following calculation is 0.62~\cite{allport2019}.
All samples were not biased during irradiation and kept at room temperature. 
Contrary to the neutron irradiation, the proton irradiation was non-uniform with
an approximately Gaussian beam profile with a FWHM between 12.5 and
\SI{15}{\milli\meter}.
In addition to the aluminum foils for dosimetry, several beam position 
monitors (BPMs) were installed in the IRRAD facility, which can be used to 
reconstruct the beam profile in horizontal and vertical direction orthogonal 
to the beam. Using this information and the aluminum foils for normalisation 
the total delivered proton fluence and the fluence profile for the modules can be estimated.
For correct positioning of the profile with respect to the module,
the position of the minimum in hit efficiency 
is set equal to the position of maximum fluence. An example is shown in Fig.~\ref{fig::beamspFit}.
The fluences $\Phi_{\text{eq}}$ in the beam spot area are about $2.4$ and 
\SI{5.4e15}{\per\square\centi\meter}, the respective numbers are quoted in the legends of Figs.~\ref{fig::effNeutr}-\ref{fig::effDefvsmax}. For the sensors bump bonded 
to the ROC4Sens readout chip, the fluences, efficiencies, and signal-to-noise ratios 
are quoted for a circular region with \SI{2}{\milli\meter} radius around the 
point of highest irradiation. The uncertainties on the fluences are estimated to 
be 17\%. 
For the sensors bump bonded to the RD53A 
readout chip, the fluences are averaged over the area of the sensor 
read out by the linear front-end, which is 
about \SI{65}{\square\milli\meter}. 

Except for the irradiation, transport and handling, the sensors are stored at 
\SI{-28}{\celsius} to avoid annealing. The integrated annealing of these steps accounts to 2-3 days at room temperature, and it is not comparable to planned annealing steps in the detector, usually 2-4 weeks long. 
  
\subsection{I-V after irradiation}
  \label{sec::sample::iv}
\begin{figure}[!htb]
  \includegraphics[width=0.99\linewidth]{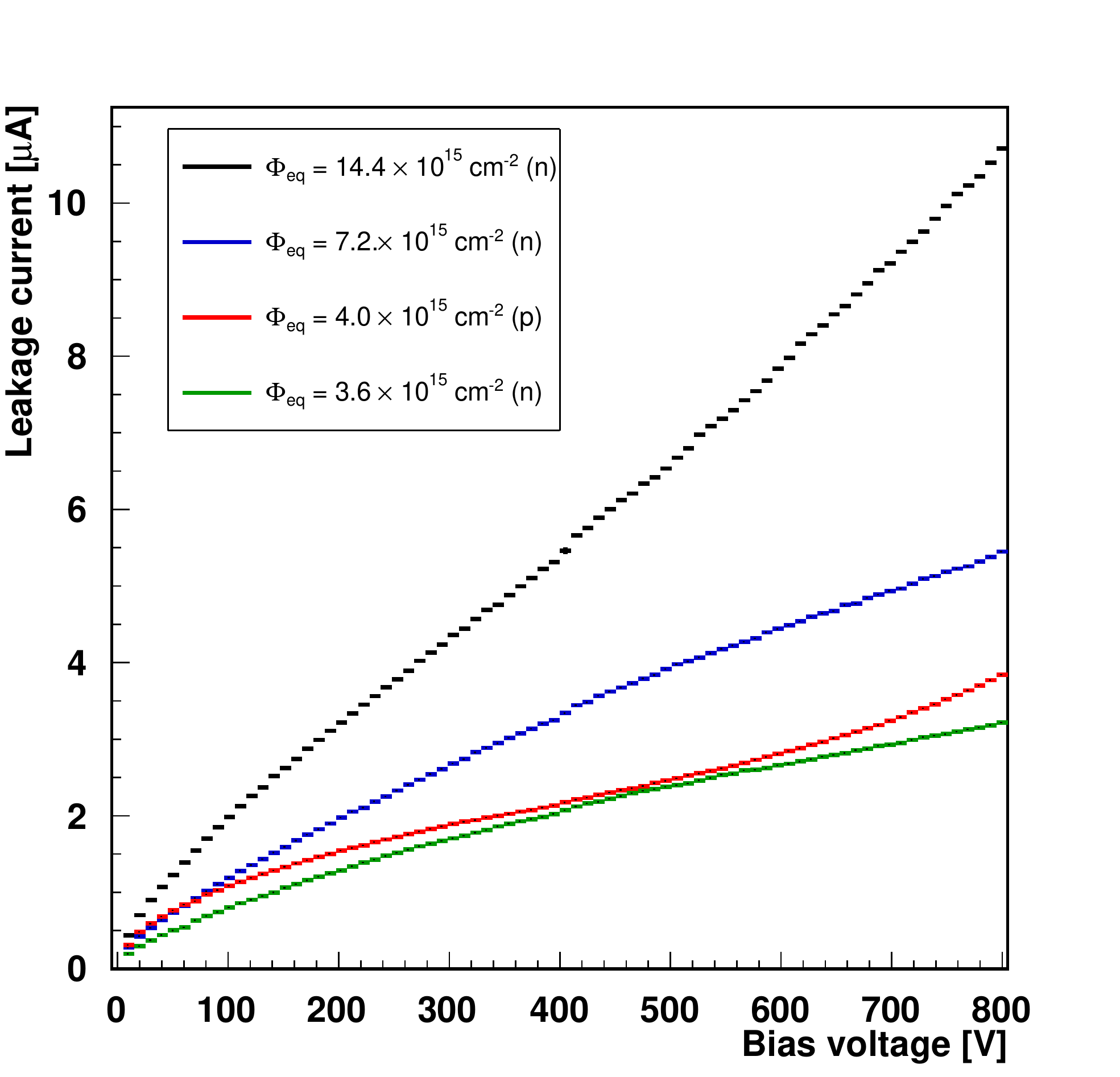}
    \caption{Leakage current as a function of the bias voltage for four different 
    ROC4Sens modules after neutron~(n) and 
    proton~(p) irradiation at \SI{-37}{\celsius}.
    The sensor area is \SI{1}{\square\centi\meter}.}
      \label{fig::IV_irr}
\end{figure}

The leakage current as a function of the bias voltage was measured during the 
beam test and in the lab. Figure~\ref{fig::IV_irr} shows the $I$--$V$ curves 
of different ROC4Sens modules irradiated with 
neutrons or protons, measured at \SI{-37}{\celsius}.
As expected, the leakage current increased with fluence. 
However, as none of the $I$--$V$ curves shows saturation it is questionable how
to extract the current-related damage factor~\cite{Moll:2018bf}. Therefore, we refrain from presenting values of this parameter, instead we discuss values of current at a fixed voltage.

For the lowest fluence the $I$--$V$ curve of the sample irradiated with neutrons 
is in good agreement with the $I$--$V$ curve of the sample irradiated with protons,
which shows that the non-ionizing energy loss scaling for the current applies. The 
NIEL hypothesis assumes that radiation damage effects scale linearly with NIEL irrespective 
of the distribution of the primary displacements over energy and space~\cite{Moll:2018bf}.
To estimate the power dissipation at a temperature of
\SI{-20}{\celsius} the current can be scaled using
$I(T)\propto T^2e^{-E_\text{a}/k_\text{B} T}$ with the activation energy $E_\text{a} = $ \SI{0.605}{\eV} 
and $k_\text{B}$ being the Boltzmann constant~\cite{Chiling2013}. For a fluence of 
$\Phi_{\text{eq}} =$ \SI{1.44e16}{\per\square\centi\meter} the
leakage current is expected to be \SI{68}{\micro\ampere\per\square\centi\meter} 
at \SI{600}{\volt} and the dissipated power is expected to be
\SI{40}{\milli\watt\per\square\centi\meter}. 
It should be noted that this leakage current value is higher compared to the requirement in Table 1, but it is obtained for a fluence much higher than specified.

\section{Beam test setup}
  \label{sec::beamttest}
The beam test measurements were performed at the
DESY~II test beam facility~\cite{Diener:2019bd} in the period
2017-2019. DESY~II provides an electron beam with
momenta between 1 and \SI{6}{\giga\electronvolt\per c}, which
is generated via a two-fold conversion and with momentum selection by 
a spectrometer dipole magnet.
For the following measurements a beam momentum of \SI{5.2}{\giga\electronvolt\per c} was 
used.
 
\subsection{Beam telescope} 
  \label{sec::beamttest:tele}
\begin{figure}[htb]
  \center
  \includegraphics[width=1\linewidth]{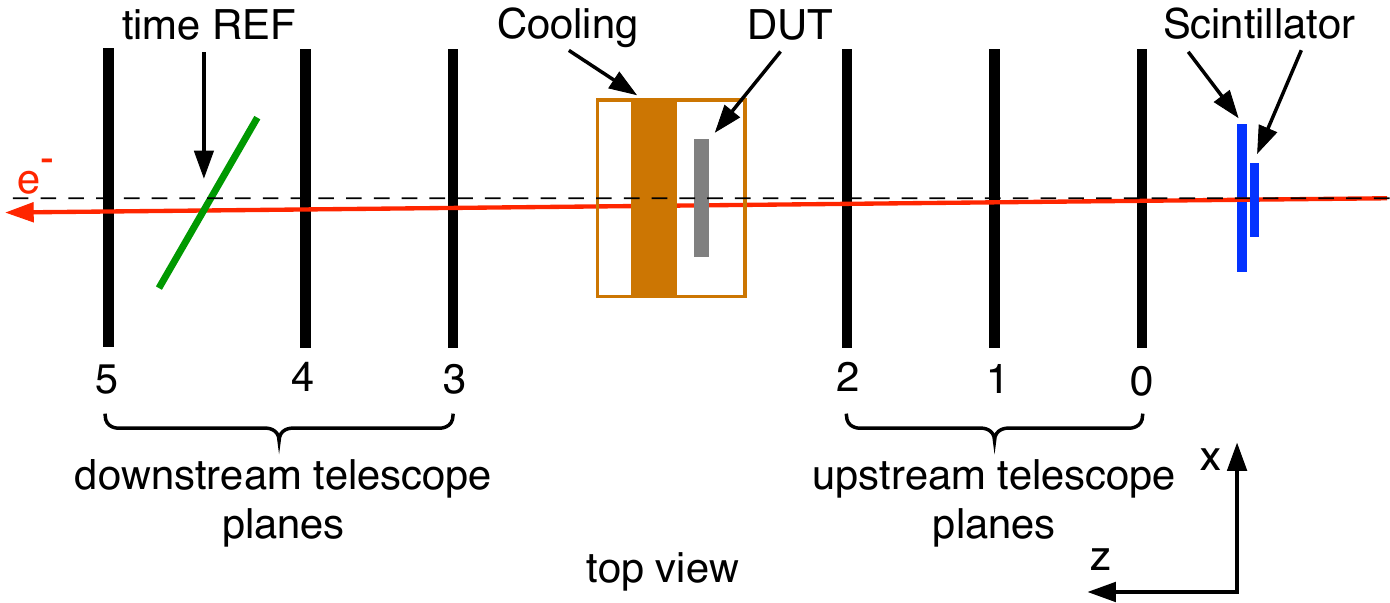}
  \caption{Sketch of the setup used for the test beam measurements, seen from the top. The time reference plane is labeled "time REF", and DUT indicates the device under test.}
  \label{fig:setup}
\end{figure}

The EUDET DATURA beam telescope~\cite{Jansen:2016ch} installed
in the beam line TB21 was used. 
The telescope consists of six planes, each equipped
with MAPS-type MIMOSA26 sensors which have a pixel size of
\num{18.4x18.4}~\si{\square\micro\meter} and are thinned
down to a physical thickness of \SI{50}{\micro\meter}. 
As shown in Fig.~\ref{fig:setup} the
planes are separated into three-plane triplets upstream and 
downstream with respect to the position of the device under test (DUT).
Operating the MIMOSA26 planes with threshold set six times the RMS noise 
an intrinsic hit resolution of a single plane of
\SI{3.24}{\micro\meter} can be achieved. Due to the long integration time of
\SI{115.2}{\micro\second} for the MIMOSA26 planes,
tracks in-time with the readout cycle of the DUT are selected 
with a CMS Phase-1 pixel module~\cite{CMSPix1_2021}, serving as time reference plane with a time tagging capability of 
\SI{25}{\nano\second}. 
Trigger scintillators upstream of the beam telescope provide a
trigger signal for the telescope, the CMS Phase-1 pixel module and the DUT.

\subsection{Pixel sensor assembly and cooling}
The pixel sensor assembly and cooling are similar to those already
used for previous CMS Phase-1 test beam measurements~\cite{Dragicevic:2017pt}.
The investigated pixel module is glued on a PCB carrier board 
with edge connectors. This carrier board is attached to a readout card
mounted on a copper plate and connected to the readout electronics.
To reduce the material in the beam,
the copper plate has a cut-out around the position of the DUT.
Inside the plate, the coolant liquid from an ethanol-based chiller circulates 
through a cooling loop to control the temperature of the DUT.
In addition, two Peltier elements operating at 5 to \SI{7}{\watt} 
in direct contact with the PCB holding the DUT are used to improve the cooling.
For thermal insulation and to prevent condensation, the copper support 
structure is placed in a plastic box, referred to as cold box, 
wrapped with ArmaFlex insulation and flushed with dry air.
The cold box is mounted on a set of two translation stages 
and one rotation stage, which allows remotely controlled movements 
in the $x$ and $y$-directions (orthogonal to the beam axis) 
and rotation around one axis of choice.

To limit the leakage current for the irradiated sensors,
the modules are cooled to \SI{-24}{\degreeCelsius} for
the setup with the ROC4Sens modules and \SI{-26}{\degreeCelsius} for the setup with the RD53A readout chip. The small difference is due to different thermal connections in the two cooling boxes used.  
Cold operation is especially important for the ROC4Sens modules since this chip has no leakage-current compensation and it has been found that already a leakage current of \SI{1}{\nano\ampere} per pixel is sufficient to significantly reduce the resistance of the feedback transistor of the preamplifier~\cite{Wiederkehr:2018mox}.

\subsection{Sensor readout and data acquisition}
A coincidence trigger is generated from the signals of two scintillators, 
read out by photomultiplier tubes (PMTs). 
To define an acceptance window slightly bigger 
than the active region of the ROC4Sens module, 
two trigger scintillators in a cross
configuration are placed upstream of the beam telescope. The output signals of the two PMTs are passed to the trigger logic unit (TLU).
The TLU is configured to send out a NIM level
trigger signal on a coincidence of the two scintillator signals.
This trigger signal is fed to a NIM discriminator to suppress occasional double pulses by choosing a sufficiently long gate. The discriminated signal is converted 
to TTL standard, split using a fanout and passed to the DTBs for the
DUT and the time reference plane. To optimise the efficiency of the time
reference plane, its trigger signal needs an additional delay of several nanoseconds.
The internal delays of the electronic devices on the trigger line 
accumulate to about \SI{112}{\nano\second}. 
 The total delay including cables
corresponds to approximately \SI{250}{\nano\second}. 
Therefore the pulse shape of the single 
pixels in the ROC4Sens modules is delayed to peak around the latter value.

\section{Data analysis}
  \label{sec::data}
In the following, only the data analysis for beam tests 
with the ROC4Sens modules as DUT is described in detail. Only one result with RD53A readout is included, and merely for completeness. A description of the tuning procedure for the RD53A readout chip is beyond the scope of this paper. 

\subsection{Online analysis}
\begin{figure}[htb]
  \centering
  \includegraphics[width=0.8\linewidth]{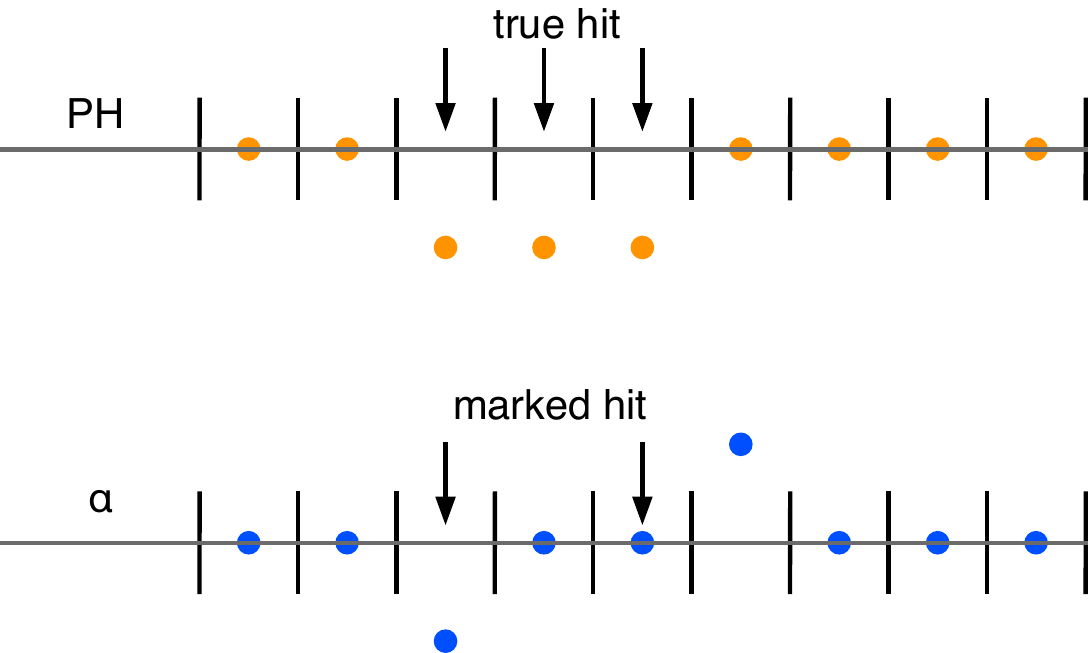}
  \caption{Hit pattern in pulse hight $PH$ and the significance $\alpha$
  for three hit pixels. The pixels marked as hit are identified by the conditions 
  given in Eq.~\ref{eq:hit}.}
  \label{fig:dph_sketch}
\end{figure}

As the ROC4Sens chip has no zero suppression, all \num{24800} pixels are 
read out for each event by the DTB and the digitised response is sent to a PC. 
To reduce the amount of stored data, only the information of possibly hit 
pixels and pixels from a region of interest (ROI) around them is stored.
This is done by applying the following procedure~\cite{Feindt:2021}:
\begin{enumerate}
\item Pedestal correction for each pixel: the pedestal is first calculated as the average response of a pixel using the first 200 events of a run. 
Subsequently, it is updated as running average.
\item Correction for baseline oscillations common to all pixels (common-mode correction): for this the
differential pulse height, $\Delta PH_{ij}$, defined as
\begin{linenomath*}
\begin{equation}
  \Delta PH_{ij} =  PH_{ij} - PH_{i-1j},
\end{equation}
\end{linenomath*}
where $PH_{ij}$ is the pedestal corrected pulse height, measured in ADC counts,
of the pixel with column and row indices $i$ and $j$, respectively, is used.
This correction can be applied in a column-wise or row-wise
sequence. Both procedures were used for the later measurements.
\item Finally, to select hits the time-dependent quantity (significance)
 \begin{linenomath*}
 \begin{equation}
   \alpha_{ij} =  \frac{\Delta PH_{ij}}{\text{RMS}(\Delta PH_{ij})} 
\end{equation}
\end{linenomath*}
is introduced as discriminator. Using a threshold
$th_\text{roi}$ a pixel $i$, $j$ is marked as hit if:
\begin{linenomath*}
\begin{equation}
   \alpha_{ij} < - th_\text{roi} \; \text{or} \; \alpha_{i+1j} > th_\text{roi}.
   \label{eq:hit}
\end{equation}
\end{linenomath*}
The usage of $\alpha$ instead of $\Delta PH$ is advantageous,
as effects of gain variations are mitigated and noisy pixels are automatically suppressed. 
The two conditions are needed to deal with clusters of hit pixels, especially if several 
consecutively read out pixels are hit. Figure~\ref{fig:dph_sketch} shows schematically a
hit pattern of three hit pixels in $PH$ and $\alpha$. It is clear that the conditions of Eq.~\ref{eq:hit} 
identify the leading and trailing hit of a cluster.

\item The pulse heights are stored for a region of interest, which consists of \num{7x7} pixels centered around a hit pixel. 
\end{enumerate}
As a compromise between efficiency of the hit identification, purity of the data sample and required disk space, all measurements were performed with $th_{\text{roi}}\approx 4$.
 
For the six MIMOSA26 sensors, the threshold is applied on the chip and 
only the positions of the pixels exceeding the threshold are stored (binary 
readout). A threshold of 5 or 6 times the individual pixel noise is used. 

For the CMS Phase-1 module used as time reference (time REF), the response of pixels above 
a threshold of 1500~e$^-$ is digitised with 8~bit 
precision and stored together with the pixel positions.

\subsection{Offline reconstruction and alignment}
\label{data:offline} 
A fast and flexible custom reconstruction and analysis software is used.
The reconstruction is performed in two steps. In the first step the reference 
tracks of the telescope are reconstructed and the telescope planes are 
aligned. In the second step the reference tracks are matched 
to the DUT and to the time reference module. Their projected track positions are matched to hits on those modules and their alignment is determined.
In both steps, an iterative approach is used, starting with loose cuts, 
still leaving a lot of combinatorial background, and iteratively using tighter cuts, resulting in a more precise alignment.
 
The alignment of the telescope starts with the readout of the binary pixel hit
information from the MIMOSA26 sensors, where noisy pixels are detected and 
removed from further analysis. Afterwards, a topological cluster algorithm is 
applied, which combines adjacent hit pixels into a cluster and calculates its 
position in local coordinates as a weighted sum of the pixel positions with the number of 
neighbouring hit pixels as weights. Fixing the position 
of plane~1 for the upstream arm and plane~4 for the downstream arm of the 
telescope allows the calculation of cluster correlation histograms and profiles 
between the planes~0 and~2 with plane~1 and planes~3 and 5 with plane~4 
to determine relative shifts in $x$, $y$ and rotations around the $z$-axis.  

Next, a triplet method is applied separately to the upstream and 
downstream arm to find initial track candidates. 
In case of the upstream arm, for all possible hits in plane~0 straight 
lines connecting to all possible hits in plane~3 are formed.
To reduce the number of combinations, track candidates with an absolute 
slope greater than \SI{5}{\milli\radian} 
are rejected. Remaining tracks are rejected, 
if no matching hit is found in plane~1 within 
\SI{50}{\micro\meter} in $x$ and $y$ of the interpolated line.
The track candidates for the downstream arm are calculated using the same method. 
The slope of the upstream and downstream triplets is used to align 
the $z$-position of planes 2 and 5.
Finally the upstream and downstream triplets are extrapolated to the nominal $z$-position of 
the DUT and correlated to determine the relative alignment between the upstream and downstream 
triplets. Only tracks for which the residuals of the $x$ and $y$ positions at the DUT between 
the two extrapolated triplets are smaller than \SI{100}{\micro\meter} are considered for the 
alignment.

The second step starts with the reconstruction of the hits in the DUT
and the time REF. For the DUT \num{7x7} pixel ROIs, which might overlap,
are read out and a fixed threshold $th_{\mathrm{pix}}$, whose value is optimised for the 
spatial resolution of each individual module, is applied.
For non-irradiated modules, the response is 
corrected for gain variations,
non-linearity, common-mode and cross talk, whereas for irradiated modules, due to the 
radiation effects on the calibration circuit, only common-mode and cross talk is corrected. 
For the DUT and the time reference plane, the same clustering algorithm as in 
Ref.~\cite{Dragicevic:2017pt}
is applied. Starting with a seed pixel the number
of hits in the cluster is obtained by adding neighbouring pixels 
that are above the threshold and adjacent to a pixel of the cluster. 
A new seed pixel is selected if there are still pixels above threshold
after removing the pixels of the cluster.
The cluster position is calculated with the Center-of-Gravity 
algorithm.

The alignment of the DUT and the time reference plane is carried out in a similar way 
as the alignment of the telescope. For the DUT, the residuals of the 
$x$- and $y$-coordinates are the difference between the cluster position 
reconstructed in the DUT and the average of the positions obtained by 
extrapolation from upstream triplet tracks and downstream triplet tracks 
to their intersection with the DUT plane. Small differences between upstream and downstream extrapolation are to be expected due to multiple scattering in the material traversed by the electrons. 
The extrapolated values are calculated from the intersection points
between track and DUT, taking into account the $z$-position and orientation
of the DUT. Then the intersection points are transformed into local DUT 
coordinates and the alignment parameters are determined as for the telescope, 
taking into account rotations around the $x$ and $y$ axis in addition. 
In case of the time reference plane, only the downstream triplet tracks are considered for 
the alignment. 

\subsection{Event selection and definition of observables} 
For the determination of the properties of the DUT, the tracks
have to fulfil additional requirements:
\begin{enumerate}
\item Residuals in $x$ and $y$ between the interception points of the extrapolated upstream and downstream triplet 
at the DUT must be $<$ \SI{30}{\micro\meter}.
\item For each extrapolated downstream triplet at the time reference plane the distance to 
the nearest other triplet must be $>$ \SI{600}{\micro\meter}.
\item Residuals in $x$ and $y$ between the track intersections and the cluster positions in the time reference plane must be $<$ \SI{150}{\micro\meter}. Such tracks are considered as in time with the DUT.
\item The tracks have to be inside of the sensitive area of the DUT (fiducial cuts).
\item A time difference of $<$ \SI{20}{\micro\second} between events recorded by the DUT and TLU is required to assure synchronization between them. 
\end{enumerate}

\subsubsection{Hit detection efficiency}
  \label{sec::data::hit}
The hit detection efficiency $\epsilon$ and its error $\sigma_{\epsilon}$ 
are defined as
\begin{linenomath*}
\begin{equation}\label{eq:eff}
  \epsilon = \dfrac{N_{\mathrm{hit}}}{N_\mathrm{t}} \text{ and } \sigma_\epsilon =
   \sqrt{\epsilon(1-\epsilon)/N_\mathrm{t}},
\end{equation}
\end{linenomath*}
where $N_{\mathrm{t}}$ denotes the number of in-time telescope tracks 
and $N_{\mathrm{hit}}$ the subset of those tracks matched with a hit in the DUT.
A hit is defined as a pixel fulfilling the conditions in Eq.~\ref{eq:hit} with $th_{\mathrm{roi}} = 4$.
This threshold is the same as the online threshold and this
definition ensures an approximately constant noise rate for all samples and 
conditions. To match a track with a DUT hit, the hit must be within 
a radius of \SI{200}{\micro\meter} of the track. For modules 
irradiated non-uniformly with protons, the efficiency is averaged over
the beam spot area.
 
\subsubsection{Charge}
For each of the $N_{\mathrm{hit}}$ tracks the charge of the cluster with 
the largest cluster charge within a radius of \SI{200}{\micro\meter}
around the track is stored. The signal is 
determined as the most probable value (MPV) of a Moyal distribution~\cite{moyal1955}, with two 
free parameters, MPV and width, fitted to the cluster charge distribution.
The Moyal function is chosen for single pixel distribution fits instead of a Landau distribution, due to its higher robustness in fits with low statistics.  
 
\subsubsection{Noise}
The noise of each pixel is defined by the RMS of its response in the
absence of particles. It defines the individual threshold of each pixel,
as discussed above. To calculate the signal-to-noise ratio, 
the noise is averaged over all pixels inside the area
(e.g.\ an area of \SI{2}{\milli\meter} radius for ROC4\-Sens modules irradiated with protons) 
considered for the determination of the efficiency and the signal.

\subsubsection{Spatial resolution}
    \label{sec::data::spat}
To reduce non-Gaussian tails in the residual distribution the selection for the determination of
the spatial resolution is more elaborate. A fixed
threshold $th_\text{pix}$ optimised for the resolution at the angle with the best resolution is used.
In addition the track is required to be isolated at the DUT.
This is ensured by requiring a minimum distance of the upstream triplet track 
extrapolated to the DUT to the nearest other triplet track of 
\SI{600}{\micro\meter}. If there are ambiguous combinations
of hits and tracks, only the closest pairs are considered.
In addition, there is a cut on the DUT residuals (Eq.~\ref{Eq:DUTres}) orthogonal to the investigated direction,
which depends on the sensor pitch (it is \SI{28.9}{\micro\meter} for the 
\num{50x50}~\si{\square\micro\meter} sensors), and finally a charge cut where the events with 
the 10\% highest charge are rejected to remove delta-electrons.
 
The resolution in the $x$-direction (similarly for the $y$-direction) is 
extracted from the distribution of the DUT residuals, $\Delta x_{\text{DUT}}$, defined as
\begin{linenomath*}
\begin{equation}
     \Delta x_{\text{DUT}} = x_{\text{DUT}} - x_{\text{TEL}},
     \label{Eq:DUTres}
\end{equation}
\end{linenomath*}
where $x_{\text{DUT}}$ denotes the position of a DUT cluster and $x_{\text{TEL}}$
the point of intersection of a telescope track in DUT coordinates, 
as discussed in Section~\ref{data:offline}.
To determine the width of this distribution, a 
method which respects the non-Gaussian nature of the distribution
for angles close to \SI{0}{\degree} and which is stable 
with respect to outliers,
a truncated RMS denoted as $\text{RMS}_\text{trc}(\Delta x_\text{DUT})$, is used. The
calculation of the truncated RMS is performed iteratively by 
discarding values outside of $\pm 6\cdot\text{RMS}_\text{trc}$. A
similar approach is applied to residuals $\Delta x_{\text{TEL}}$
of the telescope, where
\begin{linenomath*}
\begin{equation}
     \Delta x_{\text{TEL}} = x_{\text{utri}} - x_{\text{dtri}}
\end{equation}
\end{linenomath*}
with $x_\text{utri}$ being the $x$-coordinate of the extrapolation of the upstream triplet 
to the $z$-position of the DUT and $x_\text{dtri}$ defined similarly for the downstream triplet.
The effective telescope resolution, defined as the uncertainty of 
$\Delta x_\text{TEL}$, is given by
\begin{linenomath*}
\begin{equation}
     \sigma_{x_\text{TEL}} = \frac{\text{RMS}_{\text{trc}}(\Delta x_\text{TEL})}{2\cos\theta_{yD}},
\end{equation}
\end{linenomath*}
where $\theta_{yD}$ is the rotation angle of the DUT around the $y$-axis. The
factor 2 in the denominator results from averaging the position prediction 
of upstream and downstream telescope tracks, 
assuming that the uncertainty of these is the same.
Once the effective telescope resolution is known,
the resolution of the DUT is
\begin{linenomath*} 
\begin{equation}
\sigma_{x_\text{DUT}} = \sqrt{{\text{RMS}_\text{trc}(\Delta x_\text{DUT})}^2 -  {\sigma^2_{x_\text{TEL}}}}. 
\end{equation}
\end{linenomath*}

\section{Results}
  \label{sec::res}
\subsection{Results for non-irradiated modules}
Different non-irradiated types of pixel modules were investigated
in the test beam to compare their performance to expectations and to identify
less promising designs. As mentioned above, several sensor designs with 
polysilicon resistors showed problems already at this stage, which led to 
their exclusion from the further test program. 

\begin{figure}[htb]
   \centering
      \includegraphics[width=0.99\linewidth]{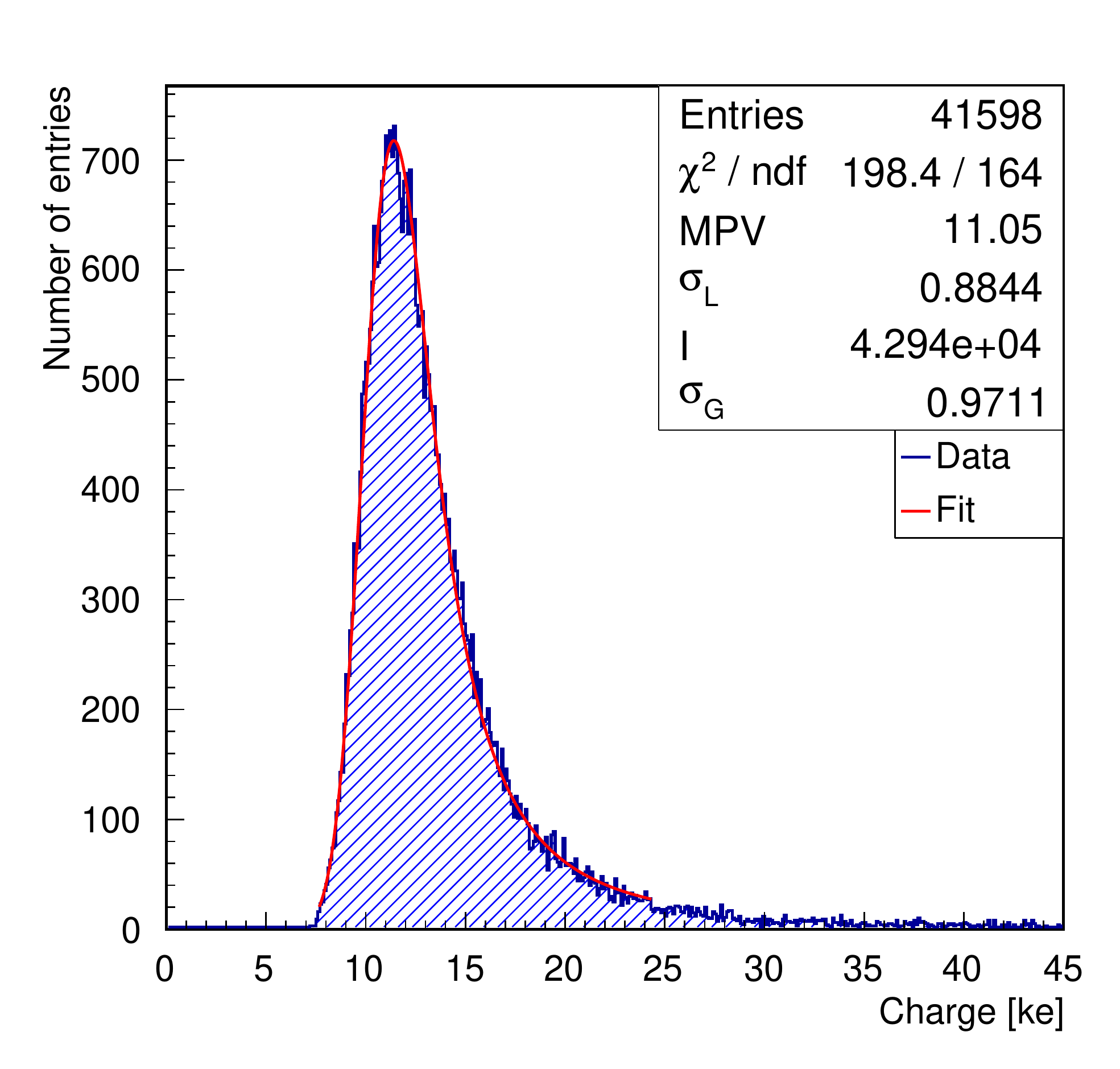}
      \caption{Cluster charge distribution measured for a non-irradiated
      sensor from a FTH150P wafer with a pixel size of 
      \num{50x50}~\si{\square\micro\meter}. 
      The measurement was performed with normal beam incidence and the sensor was biased at 
      \SI{120}{\volt}. For the fit a Landau distribution with most probable value MPV and
      width $\sigma_\text{L}$, convolved with a Gaussian distribution with width 
      $\sigma_\text{G}$, was used.}
      \label{fig::Landau50x50}
\end{figure}

In Fig.~\ref{fig::Landau50x50} a typical cluster charge distribution 
together with a fit using a Landau distribution convolved with a Gaussian distribution
is presented. The data are from a module with 
a sensor design R4S50x50-P1 which has a pixel size of 
\num{50x50}~\si{\square\micro\meter} and is from a FTH150 wafer. 
The sensor was biased at \SI{120}{\volt}. 
The measurement was done at a beam energy of 5.2~GeV with normal beam 
incidence.
For the absolute charge calibration, a gain 
calibration (pulse height vs.\ internal charge injection pulse for every pixel) was performed
and the charge was scaled by a factor of 24.3 ADC counts/ke$^-$ 
so that the most probable value is \num{11000}~e$^-$, 
which is the expected value from simulations for a sensor with
\SI{150}{\micro\meter} thickness.

\begin{figure}[!htb]
   \centering
  \includegraphics[width=0.99\linewidth]{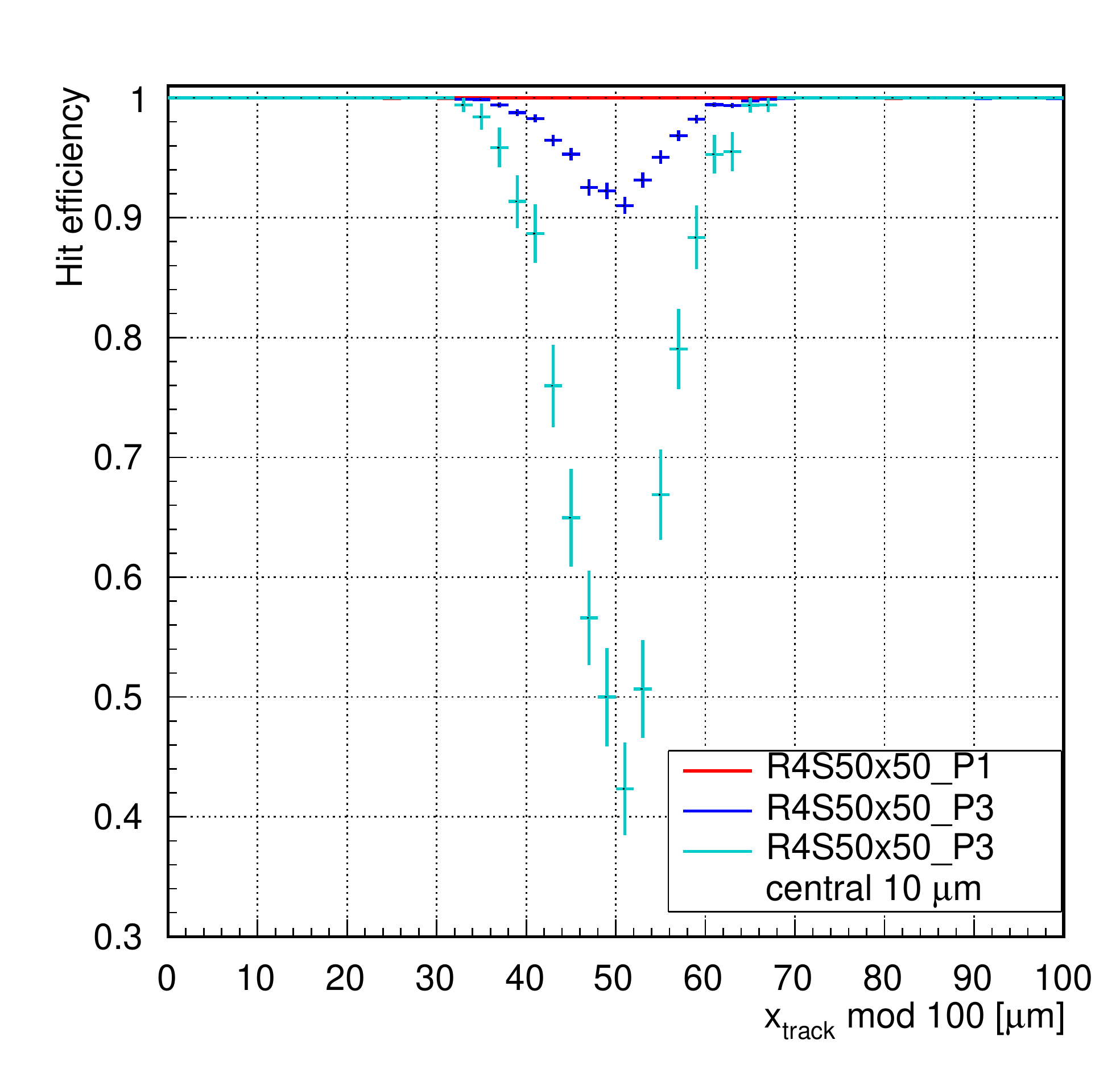}
  \caption{Projected hit efficiency vs.\ track impact point under normal incidence for two 
  non-irradiated sensors with pixel size \num{50x50}~\si{\square\micro\meter}. For the sensor with bias 
  dot (R4S50x50-P3), the projected hit efficiency drops to 92\%. The cyan curve shows the   efficiency drop in the \SI{10}{\micro\meter} region under the bias dot. 
  Considering only the central \SI{10}{\micro\meter}, in the $y$-direction, the efficiency at the bias dot drops to 
  40\%. 
  For the sensor without bias scheme (R4S50x50-P1) no significant efficiency losses are observed.}
   \label{fig::effvsxm_nonIrrad}
\end{figure}

For the non-irradiated pixel modules at a bias voltage of \SI{120}{\volt}
the hit detection efficiency is typically well above 99\%, with
the exception of the designs with bias dot. 
Significant efficiency losses are observed at the bias dot position as shown in
Fig.~\ref{fig::effvsxm_nonIrrad}, where the projected hit efficiency as a function of the in-pixel 
position is plotted for a module without bias scheme (R4S50x50-P1) and a module with
common punch-through and straight bias rail (R4S50x50-P3). For the sensor with bias dot, 
the projected hit efficiency drops to 92\%. 
The drop in a \SI{10}{\micro\meter} region in the center of the bias dot is even more severe; here the efficiency is reduced to 40\%, as shown by the cyan curve.

\begin{figure}[!htb]
   \centering
   \subfloat[]{ 
     \includegraphics[width=0.75\linewidth]{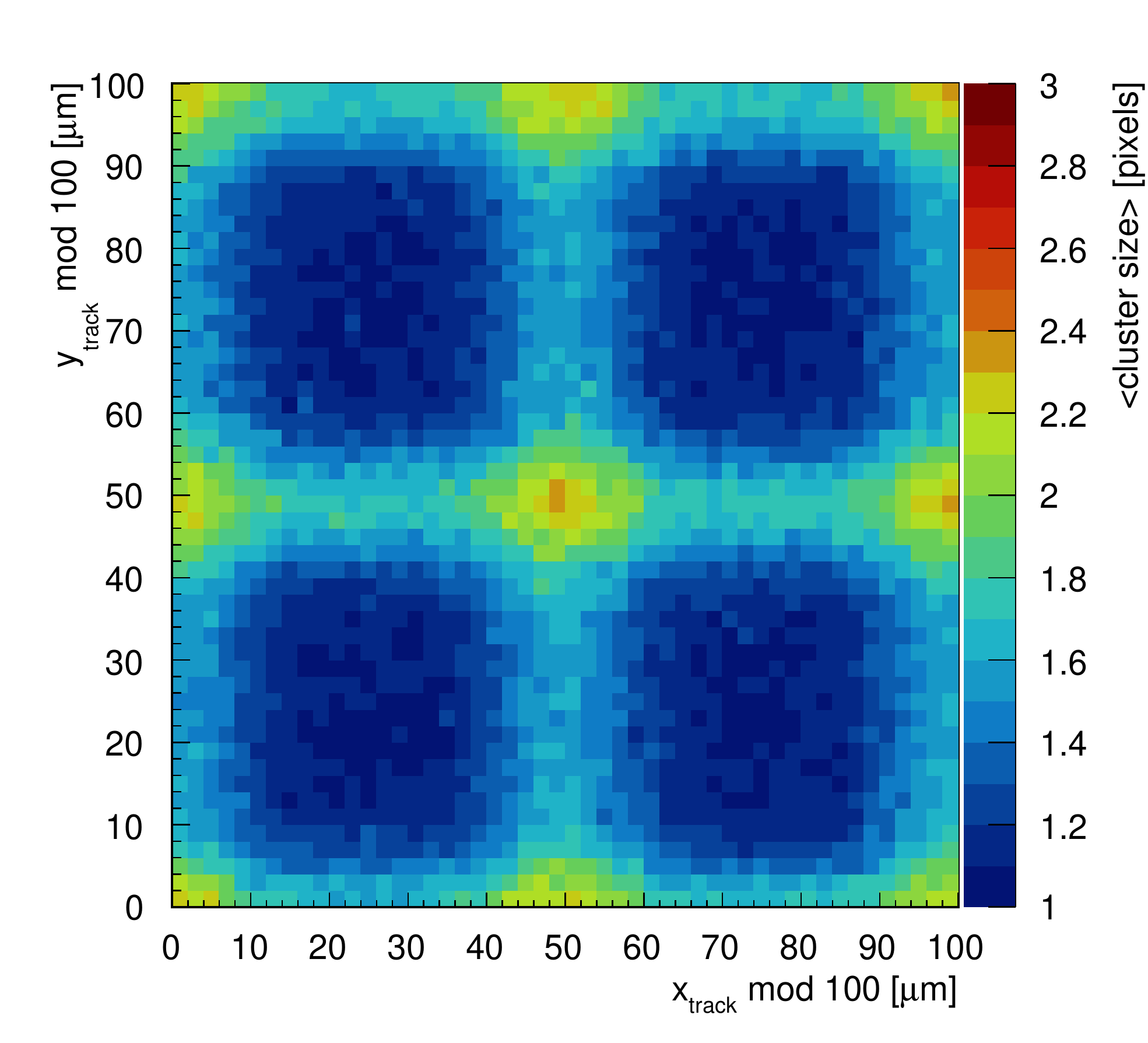}
   }\\
   \subfloat[]{
     \includegraphics[width=0.75\linewidth]{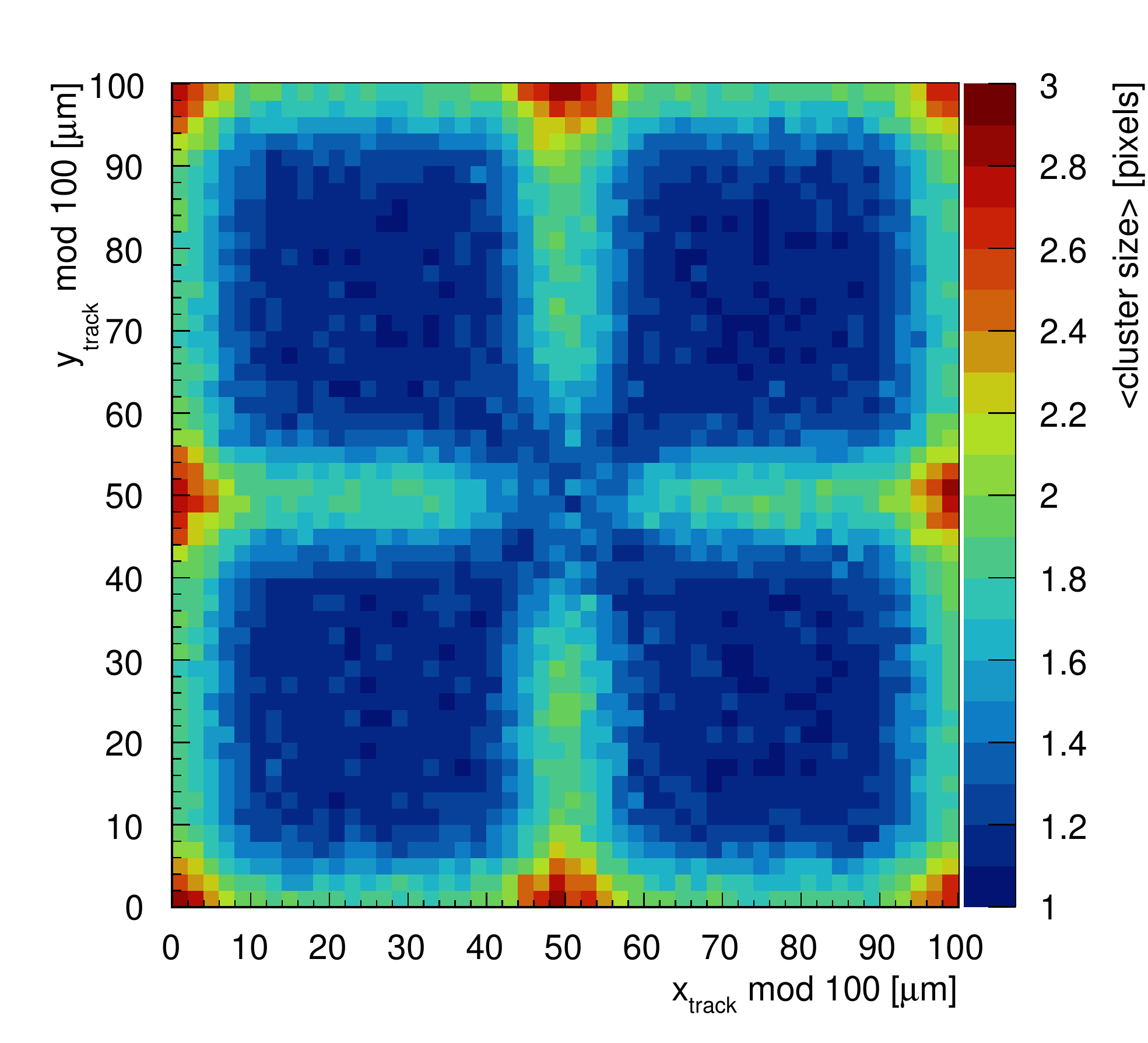}
   }
    \caption{Mean cluster size vs. track impact point under normal incidence on a 2$\times$2 pixels region for (a) a sensor without bias scheme (R4S50x50-P1) and (b) a sensor with common punch-through and straight bias rail (R4S50x50-P3).}
   \label{fig::cluster}
\end{figure}

The reduction of performance due to introduction of a bias dot is also evident
from the comparison of the mean cluster size as function of the in-pixel position
of sensors with and 
without bias dot, as shown in Fig.~\ref{fig::cluster}. 
In Fig.~\ref{fig::cluster}(a) the case without bias scheme and in
Fig.~\ref{fig::cluster}(b) the case with common punch-through and straight bias rail is presented. 
In both cases the pixel size is \num{50x50}~\si{\square\micro\meter}. 
The bias dot, which is in the centre, introduces a reduction of the cluster size.
  
\subsection{Hit detection efficiency}
   \label{sec::res::hit}
To quantify the hit detection efficiency, defined in 
Sec.~\ref{sec::data::hit}, as a function of fluence,
measurements were performed with normal beam incidence for 
voltages up to \SI{800}{\volt}. First, results after neutron 
irradiation with fluences $\Phi_{\text{eq}}$ of 
0.5, 3.6, 7.2 and \SI{14.4e15}{\per\square\centi\meter} are discussed. The 
investigated sensors are read out with the ROC4Sens readout chip.
The sensors feature a pixel size of \num{100x25}~\si{\square\micro\meter} and 
a $p$-stop pixel isolation technology, as favoured by HPK.
The pixel cell designs are without bias structure. Presented are the results of
R4S100x25-P1 shown in Fig.~\ref{fig:R4Sgds}(a) for the three 
sensors irradiated to the lower fluences, and the design R4S100x25-P7 with 
enlarged implants shown in Fig.~\ref{fig:R4Sgds}(d) for the sensor 
irradiated to the highest fluence. 

\begin{figure}[htb]
      \includegraphics[width=0.99\linewidth]{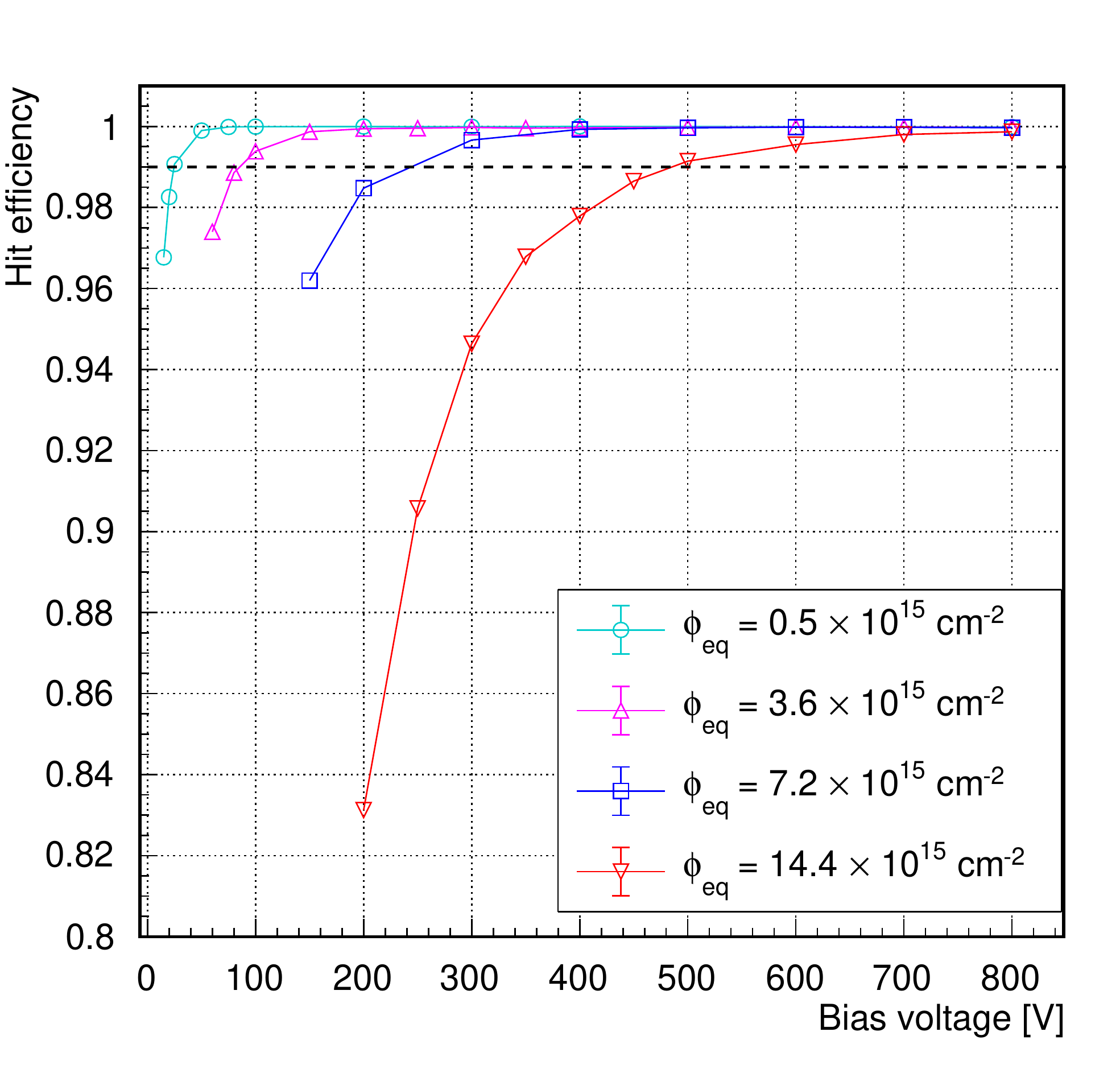}
      \caption{Hit detection efficiency after neutron irradiation
      for different fluences as a function of bias voltage.       
      The measurements were performed with vertical beam incidence angle.
      The sensors irradiated with the three lower fluences are of type R4S100x25-P1, while the sensor
      irradiated with the highest fluence is of type R4S100x25-P7. The horizontal line indicates
      a hit efficiency of 99\%. }
      \label{fig::effNeutr}
\end{figure}

In Fig.~\ref{fig::effNeutr} the hit detection efficiency
measured for the four sensors
is shown as a function of the applied bias voltage.
The required bias voltages for an efficiency of 99\%, indicated
as dashed horizontal line, are about 25, 85, 250 and \SI{500}{\volt} from the
lowest to the highest fluence, respectively. 
In general, the reason for the reduction of the hit efficiency with
increasing fluence is two-fold: due to trapping of charge carriers the signal   
decreases with increasing fluence, while the noise increases with fluence.
In addition, the electric field changes, with the region of high fields becoming 
smaller as the fluence increases.

The value of \SI{85}{\volt} for a fluence of $\Phi_{\text{eq}} =$ \SI{3.6e15}
{\per\square\centi\meter} can be compared to the full depletion voltage of below \SI{75}{\volt} before irradiation. For the highest 
fluence $\Phi_{\text{eq}} =$ \SI{14.4e15}{\per\square\centi\meter}, 
the value of \SI{500}{\volt} is well below the specified
\SI{800}{\volt}. However, even though there appears only little
difference in the amount of collected charge in strip sensors after neutron- and 
proton irradiation, as shown in Ref.~\cite{Affolder:2010ei}, 
such a conclusion must be taken with caution.

\begin{figure}[!htb]
      \subfloat[]{
      \includegraphics[width=0.99\linewidth]{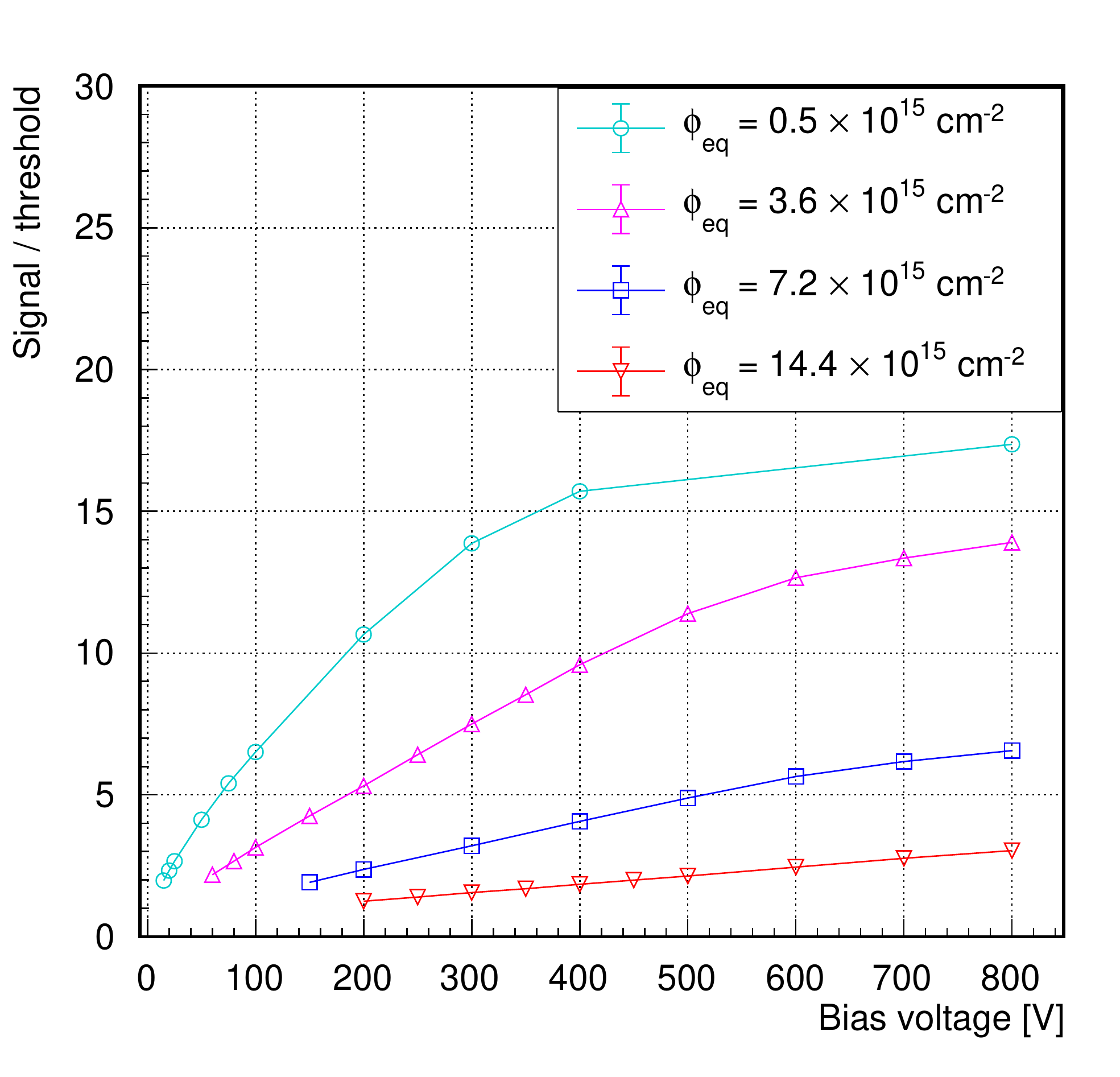}
      }\\
      \subfloat[]{
      \includegraphics[width=0.99\linewidth]{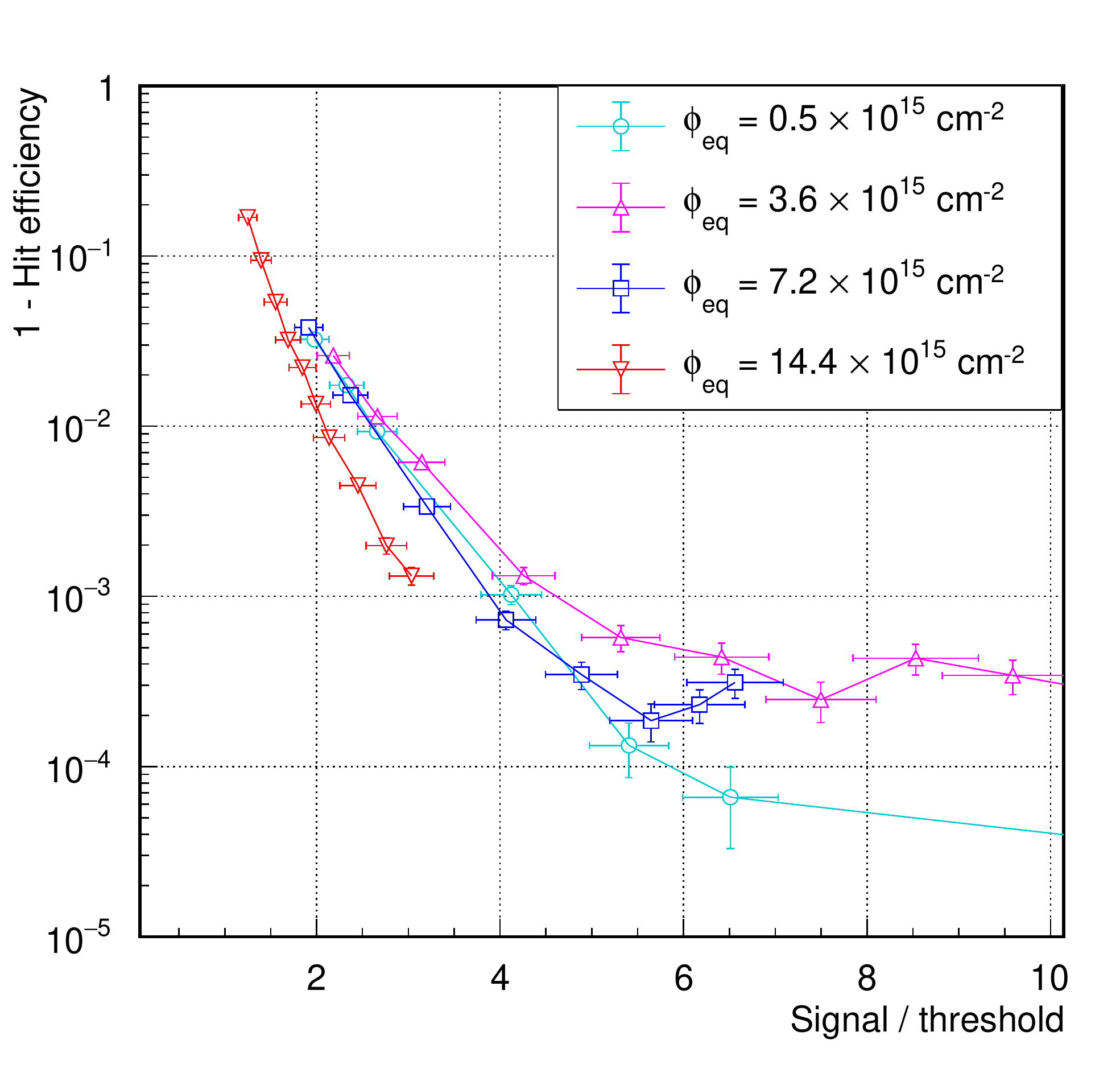}
      }
      \caption{Signal-to-threshold ratio as a function of the bias voltage (a) 
      and inefficiency as a function of the signal-to-threshold 
      ratio (b). The measurements are taken on four samples,  irradiated with neutrons
      to four fluences  $\Phi_{\text{eq}}$. All sensors have a pixel size 
      of 100$ \times$\SI{25}{\square\micro\meter} and 
      $p$-stop inter-pixel isolation. The sensors irradiated with the 
      three lower fluences are of type R4S100x25-P1, while the sensor
      irradiated with the highest fluence is of type R4S100x25-P7.}
      \label{fig::stnNeutr}
\end{figure}

In Fig.~\ref{fig::stnNeutr}(a) the signal-to-threshold ratio measured for the four
sensors is shown as a function of the applied bias voltage. 
The threshold is chosen as four times the noise --- therefore the noise rate
stays constant --- to ensure a fair comparison between the measurements taken 
under different conditions. The noise as a function of bias 
voltage is constant to within 5\%, while it doubles from 
the lowest to the highest fluence. However the variation shown in the figure
is by far dominated by the reduction of  the signal, caused by the reduction of 
collected charge.  
  
Figure~\ref{fig::stnNeutr}(b) shows the inefficiency $(1- \epsilon)$ as a
function of the signal-to-threshold ratio. 
Independently of the fluence, the three sensors of type R4S100x25-P1 reach
an inefficiency of 1\% at a signal-to-threshold 
ratio of about 2.6. This inefficiency is reached at a 
signal-to-threshold value of 2 in the case of the highest fluence. 
This is related to the larger implant of the sensor of type R4S100x25-P7, as will be shown below.
  
The mechanisms of neutron and proton radiation damage are 
known to differ at the microscopic level~\cite{Pintilie:2009qv}.
In the following, an attempt is made to quantify the different impacts on the performance of the sensors.

\begin{figure}[!htb]
		\includegraphics[width=0.99\linewidth]{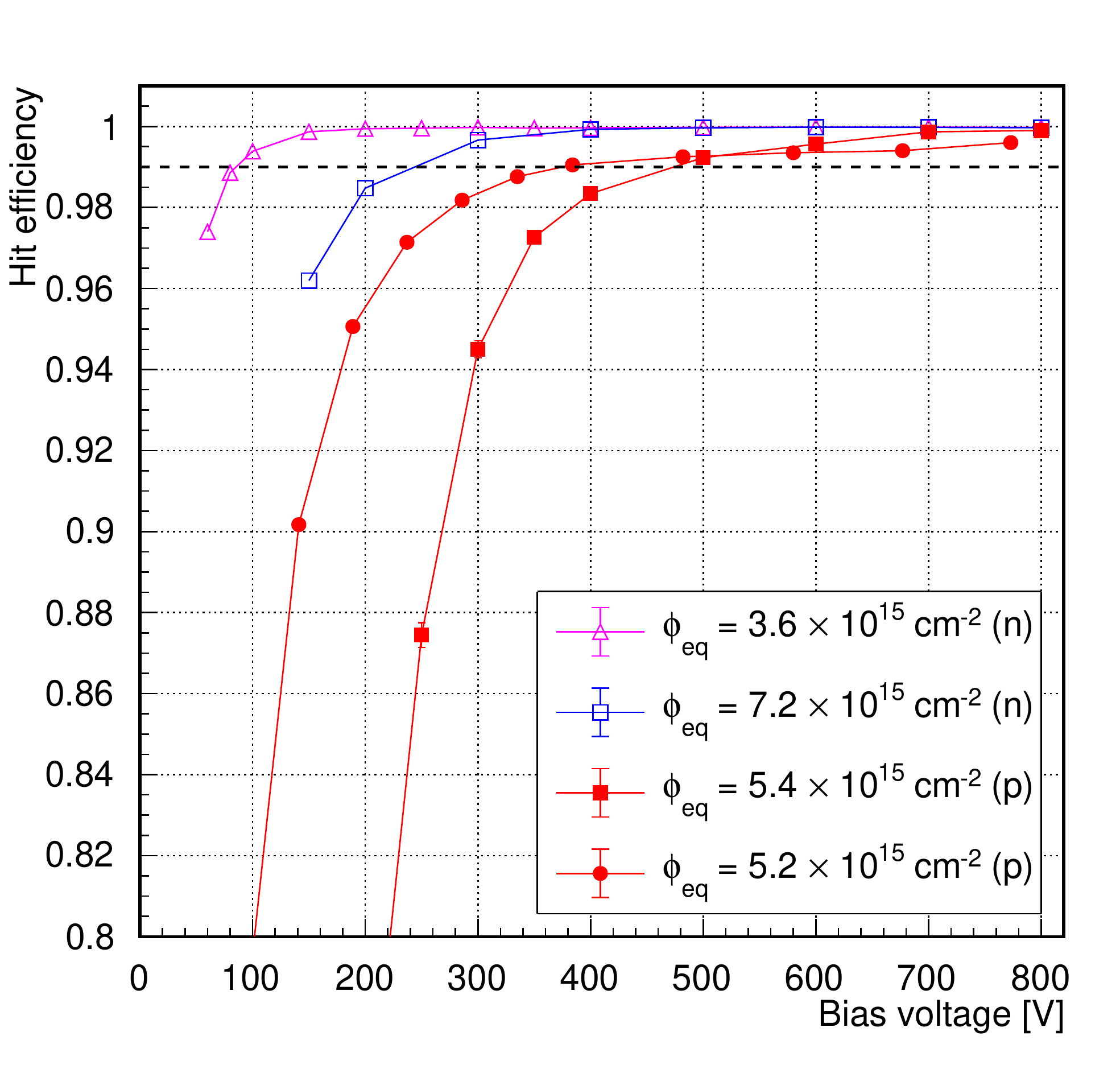}
		\caption{Hit efficiency measured at normal beam incidence 
		as a function of bias voltage for sensors irradiated with protons (p) and neutrons (n). All ROC4Sens modules are measured 
        at \SI{-24}{\degreeCelsius}. The sensors with $\Phi_{\text{eq}}=$ 5.2 and 
        \SI{5.4e15}{\per\square\centi\meter} are irradiated with protons. 
        The sensor with \SI{5.2e15}{\per\square\centi\meter}
        (red circles) is bump bonded to a RD53A chip and measured at approximately the same temperature as the ROC4Sens modules.}
		\label{fig::effNvsp}
\end{figure}

The efficiency as a function of the bias voltage for two sensors irradiated with protons to $\Phi_{\text{eq}}=$ 5.2 and \SI{5.4e15}{\per\square\centi\meter} is shown in Fig.~\ref{fig::effNvsp}. 
For comparison, the two intermediate neutron 
fluences from Fig.~\ref{fig::effNeutr} are included. It is concluded
that the modules irradiated with protons require significantly
higher operating 
voltages than those irradiated with neutrons for an efficiency of 
99\%, for which there are two reasons.
One is the higher (factor of 30) ionising dose deposited by the proton beam.
Since the ROC4Sens readout chip is more sensitive to ionising radiation, the steep  rise to about 
95\% occurs at higher bias voltages\footnote{The module with the RD53A readout chip has lower efficiency due to 0.5\% of dead pixels, which have not been excluded from the analysis.}. The second reason is the difference in bulk damage, which is investigated
in Ref.~\cite{Kramberger:2014ty} for neutron and pion irradiation.

These measurements show that the tested sensors reach an efficiency of
99\% for bias voltages significantly below \SI{800}{\volt} for a fluence of \SI{5e15}{\per\square\centi\meter}. 

\subsection{Sensor Design Comparisons}
    \label{sec::res::designs}
To choose the optimal sensor layout for the upgraded detector, 
modules with different sensor designs are compared after irradiation.
  
\begin{figure}[htb]
		\includegraphics[width=0.99\linewidth]{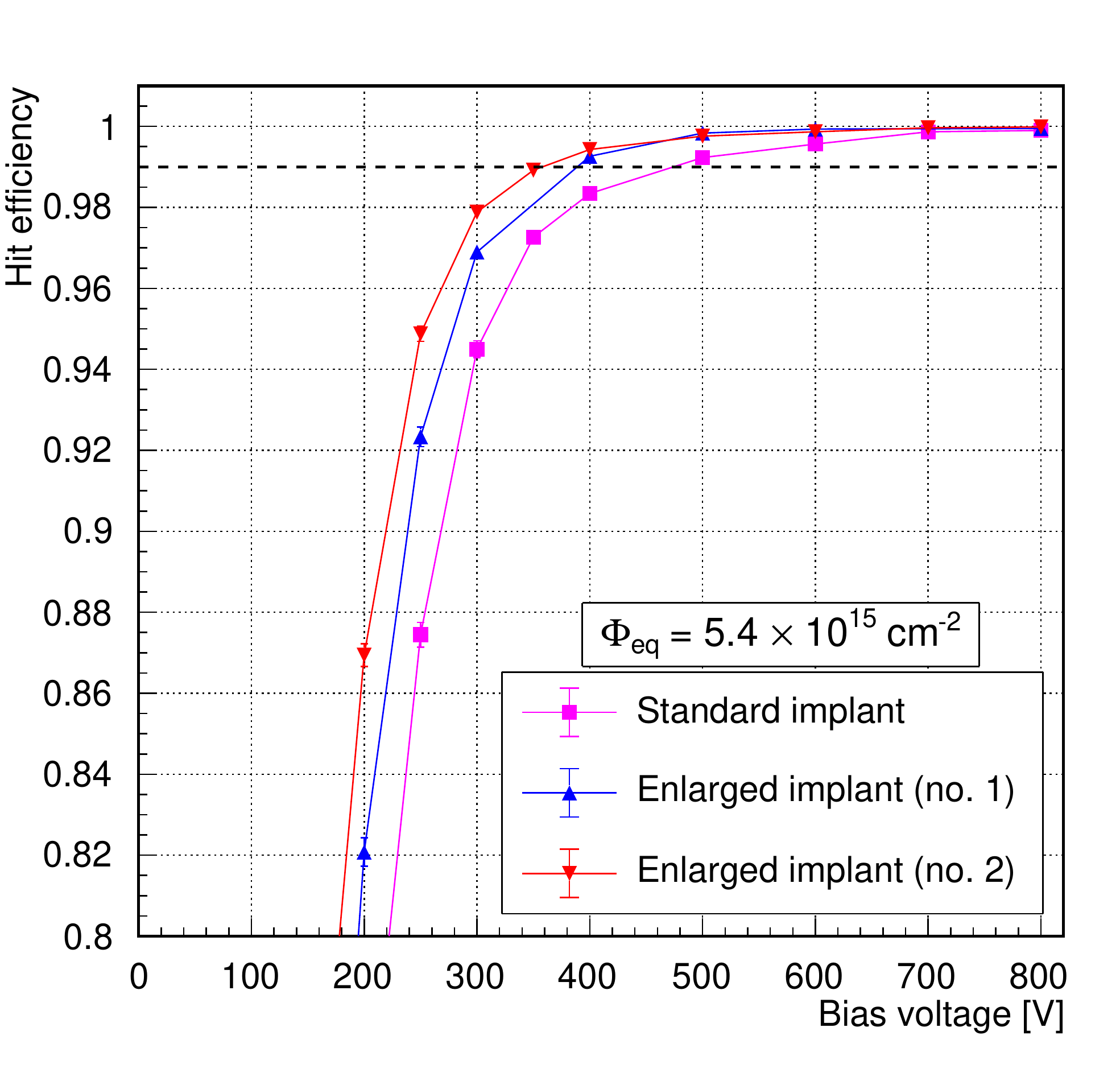}
		\caption{Hit efficiency measured at normal beam incidence 
		as a function of bias voltage for irradiated
        sensors with protons, with and without enlarged implants.}
		\label{fig::effDefvsmax}
\end{figure}

\begin{figure}[!htb]
  \begin{center}
		\includegraphics[width=0.99\linewidth]{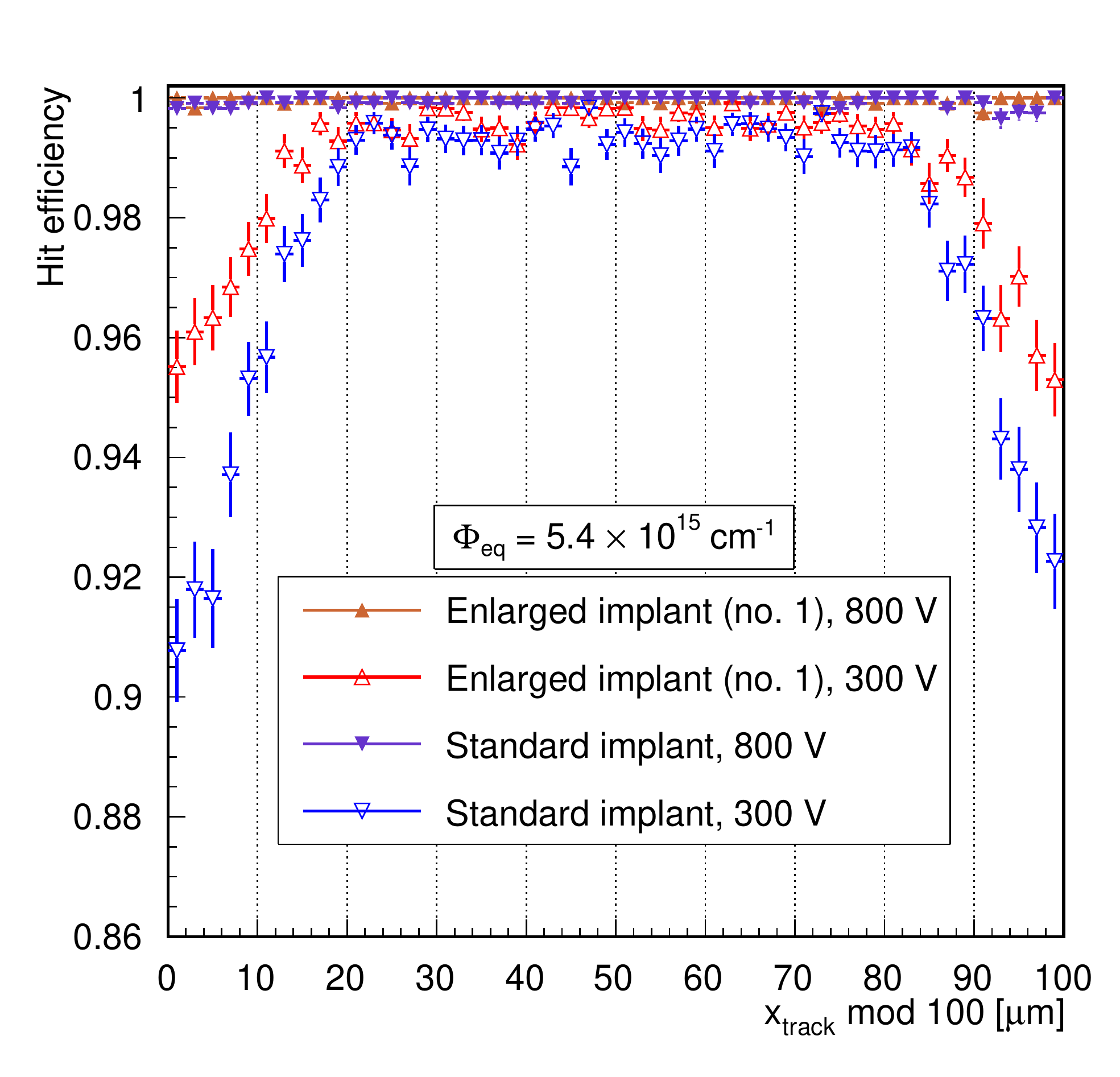}
		\caption{Hit efficiency measured at normal beam incidence as a function of the
		position inside one pixel along the \SI{100}{\micro\meter} direction.
		The sensors are the same as in Fig.~\ref{fig::effDefvsmax}, irradiated with a fluence of
		\SI{5.4e15}{\per\square\centi\meter}. The results for sensors with and without 
        enlarged implants are shown at 300 and \SI{800}{\volt}. At \SI{300}{\volt} the 
        efficiency around the pixel boundaries at 0 and \SI{100}{\micro\meter} is 
        about 3\% higher for the design with enlarged implants, while 
        the efficiencies are all compatible within 1\% in the central region.}
		\label{fig::effInpix}
	\end{center}
\end{figure}

Wider $n^+$ implants are expected to yield higher hit efficiencies~\cite{Buchanan2017}.
However, the risk of breakdown before irradiation is increased, due to the 
potentially higher field at the $p$-stop isolation.
Current-voltage measurements were performed on about 70 sensors with 
enlarged implants, and no evidence of breakdown was observed. 
In Fig.~\ref{fig::effDefvsmax} a comparison of the hit efficiency of two sensors with
and one sensor without enlarged implant is shown. 
Indeed a higher hit efficiency is observed for
the design with wider implants at the same bias voltage.
As shown in Fig.~\ref{fig::effInpix}, this is due to reduced efficiency losses
at the pixel boundaries.
Given the excellent performance of the designs with enlarged pixel implants,
this design will be further tested in the next prototyping steps.

\begin{figure}[!htb]
	\includegraphics[width=0.99\linewidth]{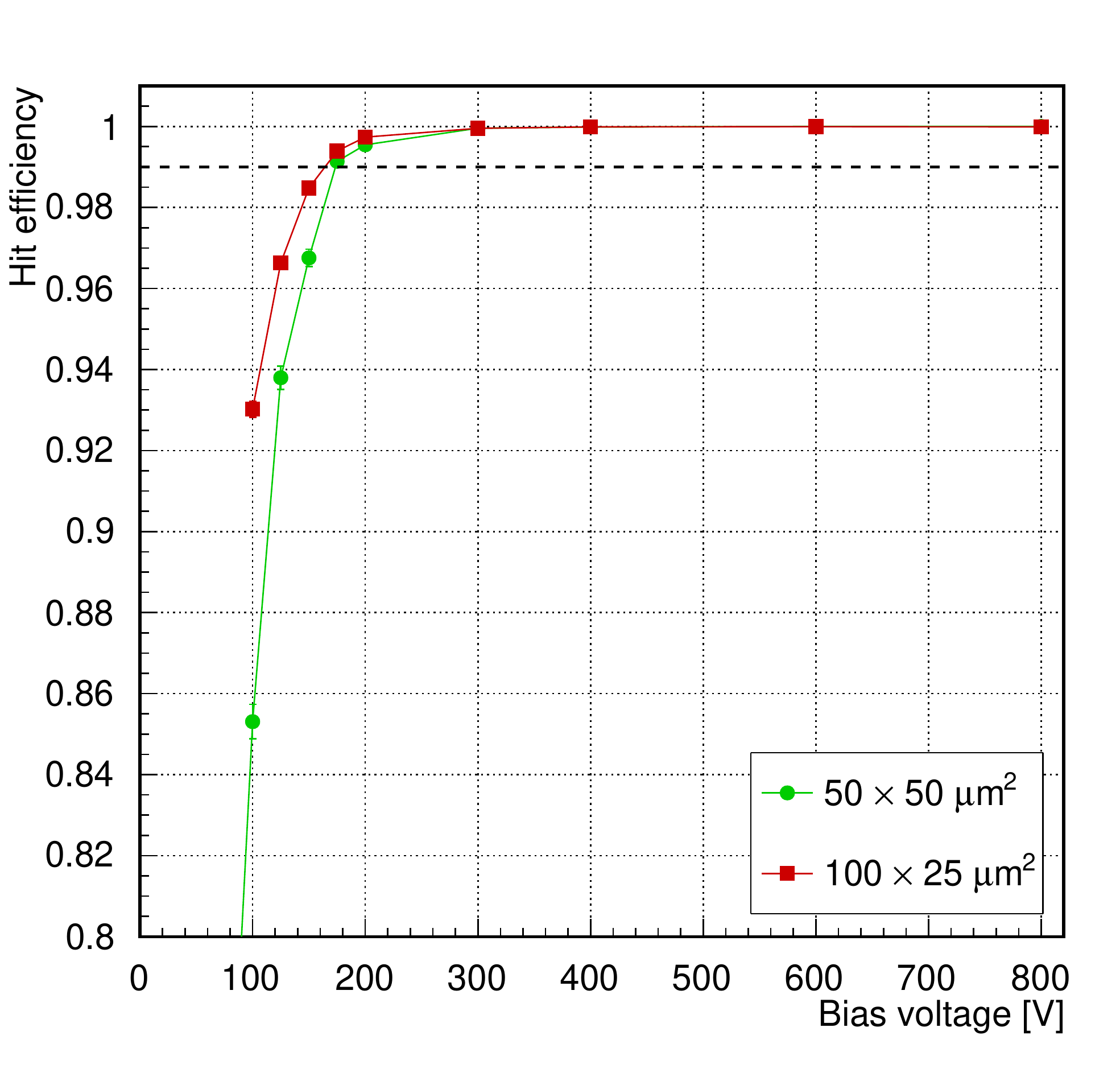}
	\caption{Hit efficiency for normal beam incidence as a function of bias 
	voltage for sensors with pixel sizes of 50$\times$\SI{50}{\square\micro\meter} (R4S50x50-P1) 
	and 100$\times$\SI{25}{\square\micro\meter} (R4S100x25-P1).
	 Both sensors were irradiated with protons to a fluence of 
	 $\Phi_{\text{eq}}=$ \SI{2.4e15}{\per\square\centi\meter}.}
	\label{fig::eff50vs25}
\end{figure}

The comparison of sensors with pixel sizes of 
\num{50x50}~\si{\square\micro\meter} and \num{100x25}~\si{\square\micro\meter} 
shown in Fig.~\ref{fig::eff50vs25} 
shows only minor differences. 

\subsection{Charge losses at the bias dot}
\begin{figure}[!htb]
	\includegraphics[width=0.99\linewidth]{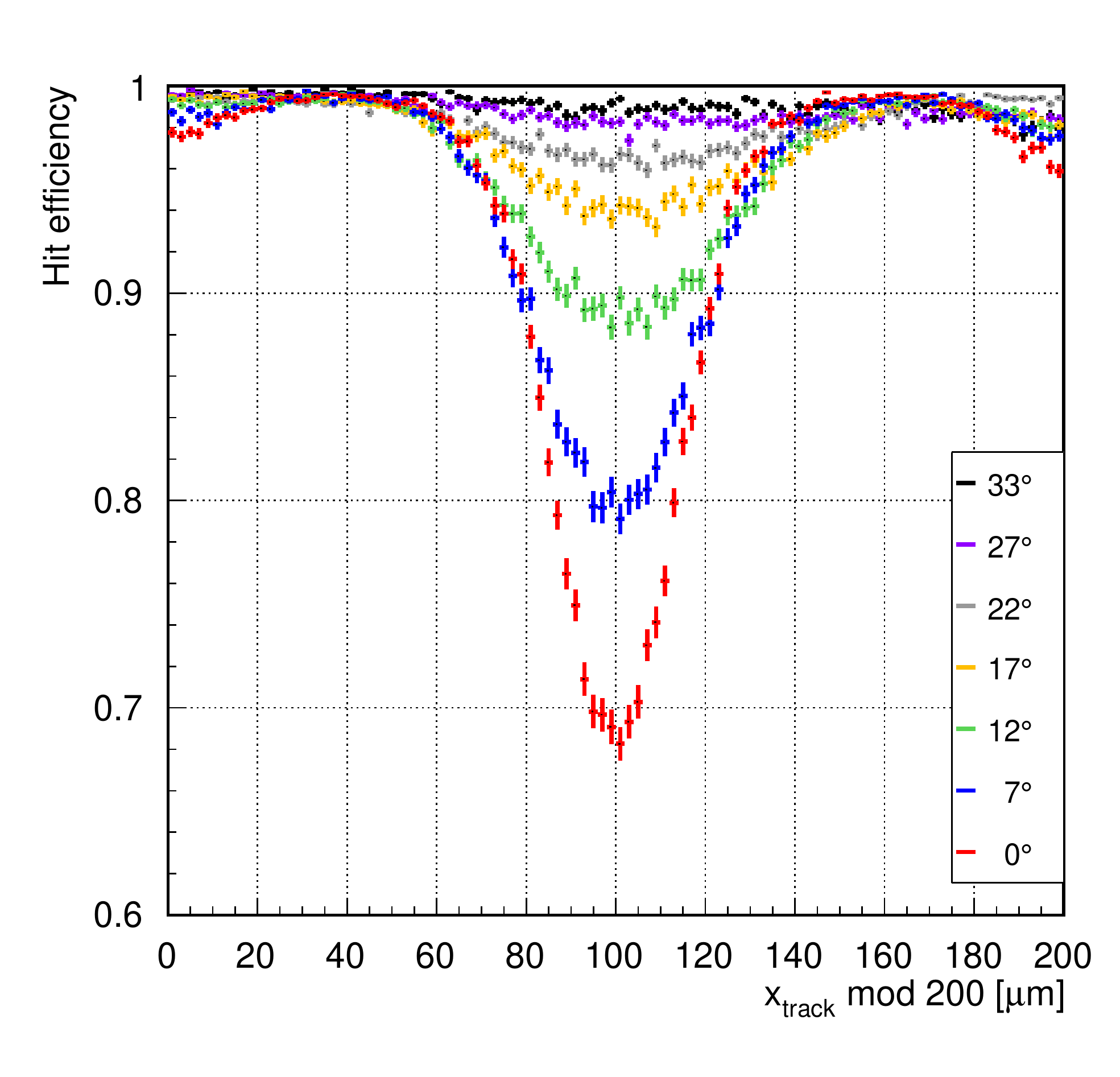}
	\caption{Hit efficiency as a function of the position inside two pixels along the
      \SI{100}{\micro\meter} direction for various track angles measured at a bias voltage of \SI{800}{\volt}. 
      The track angle is defined with respect to the perpendicular to the sensor plane.
      The inclination is in the 
      \SI{100}{\micro\meter} direction. 
      The measured sensor is of type RD53A100x25-P2 read out by a RD53A readout chip. 
      The sensor was irradiated with protons to a fluence of 
      $\Phi_{\text{eq}}=$ \SI{5.6e15}{\per\square\centi\meter}. }
	\label{fig::effBdot}
\end{figure}

For sensors with a bias dot, charge losses are expected when tracks hit the
bias dot with an angle almost perpendicular to the sensor plane. To assess these losses in 
detail, the efficiency as a function of the position in the pixel is shown in 
Fig.~\ref{fig::effBdot} for angles between 0 and \SI{33}{\degree}. The 
investigated sensor is read out by an RD53A readout chip and was irradiated
with protons to a fluence $\Phi_{\text{eq}}$ of \SI{5.6e15}{\per\square\centi\meter}.
The sensor is of type RD53A100x25-P2, shown in Fig.~\ref{fig:RD53Agds}(b).

It is observed that angles larger than \SI{22}{\degree} are needed to
overcome the efficiency loss at the bias dot, which is as high as 
30\% for \SI{0}{\degree}. Since angles close to 
\SI{0}{\degree} are expected to be frequent in the forward pixel detector,
the design without a bias dot is clearly favoured. 

\subsection{Spatial resolution}
Detailed studies of the spatial resolution after irradiation have been performed
with the DATURA telescope only for sensors with a pixel size of  
\num{50x50}~\si{\square\micro\meter}.
In the following, the measurements before irradiation and
after neutron or proton irradiation are presented as a function of
the beam incidence angle. Measurements of the non-irradiated sensor
were made at \SI{120}{\volt}, while the irradiated samples were measured at 
\SI{800}{\volt} to maximise the collected charge. 
The reconstruction of the resolution and the event selection was done
as described in Sec.~\ref{sec::data}.

\begin{figure}[!htb]
      \subfloat[]{
      \includegraphics[width=0.99\linewidth]{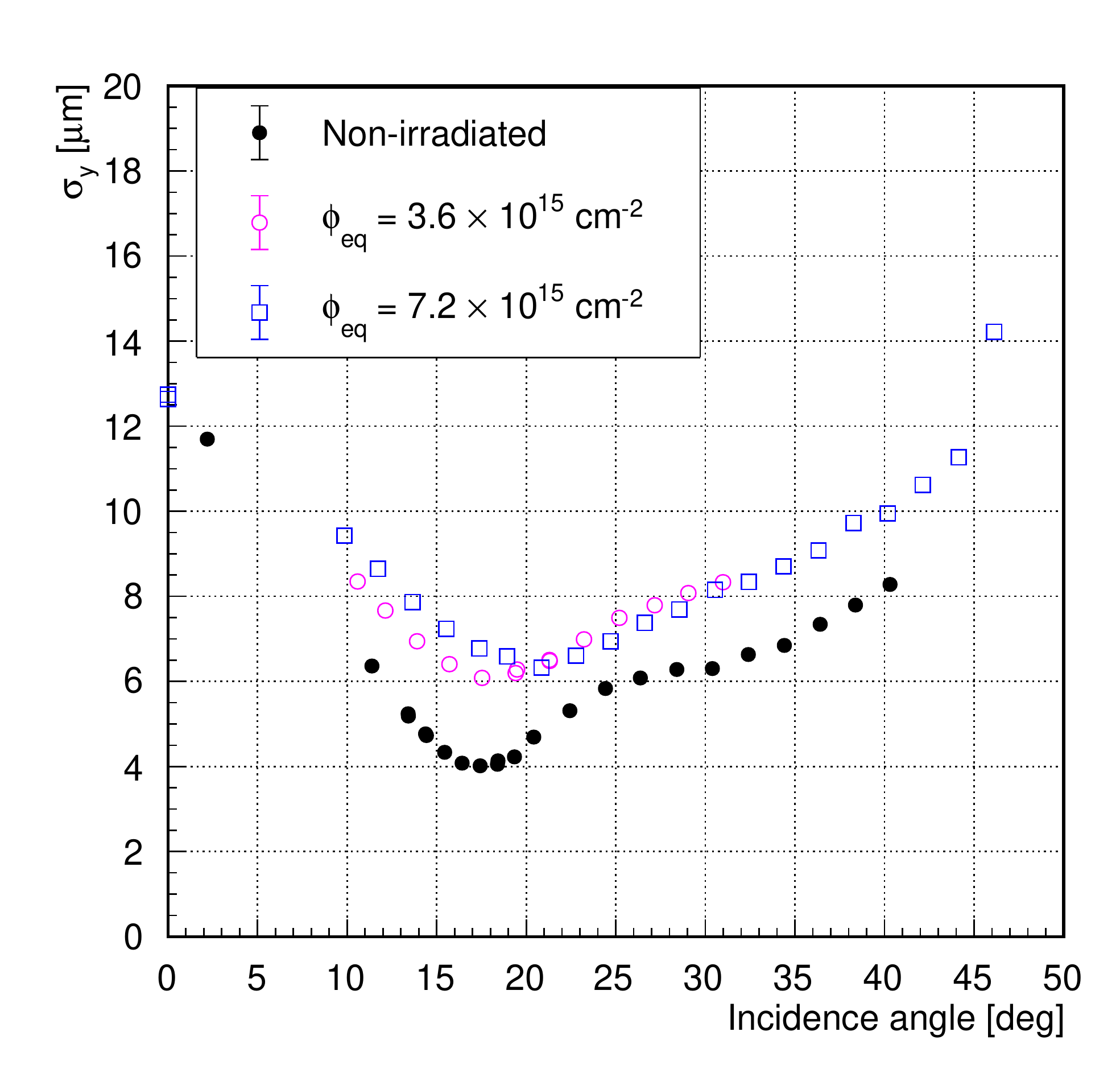}
      }\\
      \subfloat[]{
      \includegraphics[width=0.99\linewidth]{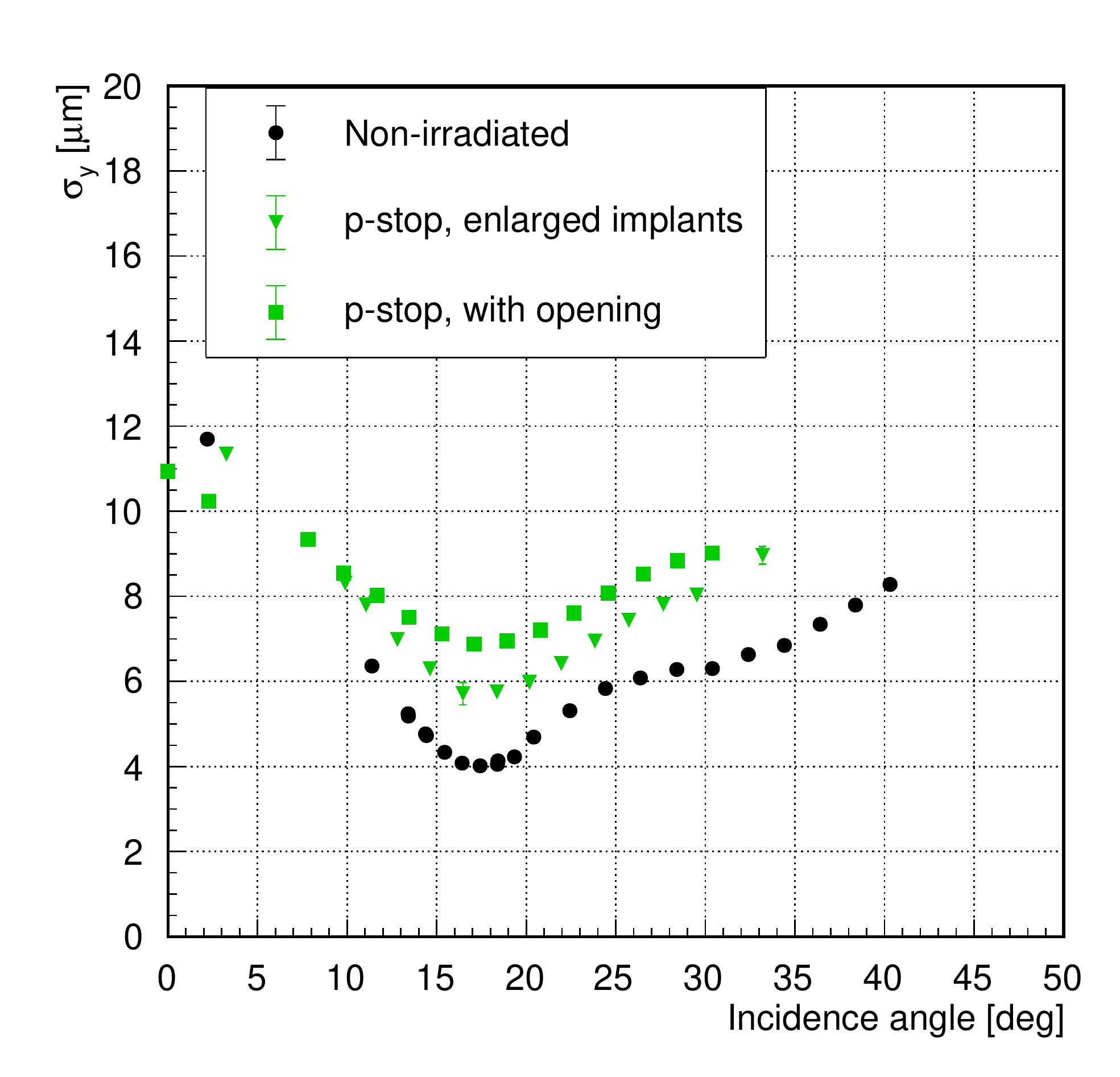}
      }
      \caption{Spatial resolution measured at 800~V as function of the track angle, 
      for (a) a non-irradiated sensor and two sensors irradiated with neutrons and (b) a 
      non-irradiated sensor and two sensors irradiated with protons to a
      fluence of \SI{2.3e15}{\per\square\centi\meter}. The investigated modules have a pixel size of 50$\times$\SI{50}{\square\micro\meter}.}
      \label{fig::resolution}
\end{figure}

Sensors irradiated with neutrons to fluences of
$\Phi_{\text{eq}} =$ \SI{3.6e15}{\per\square\centi\meter} and  
$\Phi_{\text{eq}} =$ \SI{7.2e15}{\per\square\centi\meter} have been investigated.
The studies include a non-irradiated sensor of R4S50x50-P8 type, which is 
with an enlarged implant and without bias structure, 
for comparison with the results after irradiation.
The sensor irradiated with the higher fluence is of the R4S50x50-P1 type, while
the sensor irradiated with the lower fluence is the corresponding $p$-spray 
version.
The spatial resolution in $y$ direction is studied as a function of the rotation angle around the $x$-axis, $\theta_x$.  
The analysis has been performed in two steps. 
In the first step the threshold is optimised
at the angle with best resolution (optimal angle), which is $\theta_x = $ \SI{17.5}{\degree} for
the lower fluence and $\theta_x = $ \SI{20.9}{\degree} for the higher fluence.
This has to be compared to $\theta_x = $ \SI{17.4}{\degree} for a non-irradiated sensor. 
The optimal angle for the larger fluence is significantly higher. This is due to the fact that the depth dependence of the charge collection increasingly reduces the effective thickness of the pixel sensor with increasing fluence.
The optimal threshold values are determined as 12, 18 and 20 ADC counts,
respectively, from the lowest to the highest fluence.
They correspond to signal-to-threshold values of 5\%, 
8\% and 11\% of the Landau MPV.
In the second step the spatial resolution as function of the beam incidence angle 
is determined using these threshold values. In Fig.~\ref{fig::resolution}(a) the
results are shown in comparison to those of the non-irradiated sensor. 
The shapes of the curves are qualitatively 
similar. However, the resolution at the optimal angle
degrades from \SI{4.0}{\micro\meter} to 
$6.1\pm $\num{0.1}~\si{\micro\meter} after
$\Phi_{\text{eq}}=$ \SI{3.6e15}{\per\square\centi\meter} and to
$6.3\pm $\num{0.1}~\si{\micro\meter}
after $\Phi_{\text{eq}} =$ \SI{7.2e15}{\per\square\centi\meter}.

To study the resolution after proton irradiation, two samples of different type,
irradiated with protons to nearly the same fluence of
$\Phi_{\text{eq}} =$ \SI{2.3e15}{\per\square\centi\meter}, were used.
One is of type R4S50x50-P2, which has an open $p$-stop isolation, and
one of type R4S50x50-P8, which has an enlarged implant.
As in the case of the neutron irradiation the measurements 
have been performed at 
\SI{800}{\volt}. The threshold optimisation at the optimal angle results
in 16~ADC counts for the sensor with enlarged implants and 18~ADC counts 
for the sensor with the open $p$-stop isolation, which corresponds in both cases to
10\% of the Landau MPV. In Fig.~\ref{fig::resolution}(b) the 
spatial resolution as a function of track angle determined with these 
threshold values is shown in comparison to the non-irradiated sensor. 
The resolution at the optimal angle
degrades from $4.02\pm$\num{0.03}~\si{\micro\meter} to
$5.7\pm$\num{0.3}~\si{\micro\meter} for the
design with the enlarged pixel implant and to 
$6.9\pm$\num{0.1}~\si{\micro\meter}
for the open $p$-stop after
$\Phi_{\text{eq}} = $~\SI{2.3e15}{\per\square\centi\meter}.

\section{Conclusions}
\label{sec::concl}
This paper summarizes the qualification of planar pixel sensor designs suitable for the CMS Inner Tracker, investigated using an R\&D chip (ROC4Sens).  
The results presented in this paper demonstrate that some of the designs implemented on an HPK submission
reach efficiencies of 99\% for minimum 
ionising particle tracks normal
to the sensor plane at voltages
above 500 and \SI{400}{\volt} after neutron and proton irradiation to 
fluences $\Phi_{\text{eq}}$ of up to 14.4 and \SI{5.4e15}{\per\square\centi\meter},
respectively. The higher value is above the fluence expected for planar pixel sensors in
the upgraded CMS Inner Tracker, which is about \SI{1.2e16}{\per\square\centi\meter}. 

The intrinsic single plane resolution along the \SI{50}{\micro\meter} pitch
direction is shown to be \SI{4.0}{\micro\meter} for the non-irradiated sample at the optimal angle, 
while it worsens to \SI{6.3}{\micro\meter} after
neutron irradiation of $\Phi_{\text{eq}} = $ \SI{7.2e15}{\per\square\centi\meter}.

The measurements presented in this paper have informed the choice of the sensor design, together with 
other studies such as physics performance simulations and thermal modelling. Planar sensors with a 
pixel size of \num{100x25}~\si{\square\micro\meter} will be used everywhere except in the innermost barrel layer, where 3D sensors with the same pixel size will be employed. The planar sensors will not
feature a punch-through bias dot, but an enlarged implant. A cell design similar to that of 
Fig.~\ref{fig:RD53Agds}(a) is going to be used. Parylene coating will be used for spark protection.

Further studies, including measurements at higher irradiation fluences that require a calibrated 
RD53A readout chip, are ongoing. Preliminary studies for angles up to \SI{40}{\degree} were presented in Ref.~\cite{ebrahimi2019}.

\section*{Acknowledgements}
 \label{sect:Acknowledgement}
This work was supported by the German Federal Ministry of Education
and Research (BMBF) in the framework of the "FIS-Projekt - Fortf\"uhrung des
CMS-Experiments zum Einsatz am HL-LHC: Verbesserung des Spurdetektors 
f\"ur das Phase-2 Upgrade des CMS-Experiments"
and supported by the H2020 project AIDA-2020, GA no.\ 654168.
The measurements leading to these results have been performed at the Test Beam Facility at DESY Hamburg 
(Germany), a member of the Helmholtz Association (HGF).

The tracker groups gratefully acknowledge financial support from the following funding agencies: BMWFW 
and FWF (Austria); FNRS and FWO (Belgium); CERN; MSE and CSF (Croatia); Academy of Finland, MEC, and HIP 
(Finland); CEA and CNRS/IN2P3 (France); BMBF, DFG, and HGF (Germany); GSRT (Greece); NKFIA K124850, and 
Bolyai Fellowship of the Hungarian Academy of Sciences (Hungary); DAE and DST (India); INFN (Italy); PAEC 
(Pakistan); SEIDI, CPAN, PCTI and FEDER (Spain); Swiss Funding Agencies (Switzerland); MST (Taipei); STFC 
(United Kingdom); DOE and NSF (U.S.A.). This project has received funding from the European Union's 
Horizon 2020 research and innovation programme under the Marie Sk\l odowska-Curie grant agreement No 
884104 (PSI-FELLOW-III-3i). Individuals have received support from HFRI (Greece).

\bibliographystyle{elsarticle-num-names}
\bibliography{CMS_Pixel}
\appendix
 \renewcommand*{\thesection}{\Alph{section}}
 \section{Appendix A: Sample list}
 \label{sect:AppendixA}
 \begin{table}[htb]
  \centering
  \caption{List of all single chip modules used in these studies. 
  The letters P and Y at the end of the material identifiers refer to $p$-stop and $p$-spray modules, respectively. In the fourth column, the proton irradiation at the CERN PS-IRRAD is labelled as p and the 
  neutron irradiation as n.} 
   \begin{tabular}{@{}ccccc@{}}
   \toprule
   Nr. & Mat. & Type & Irr. &$\Phi_{\text{eq}}$ \\
          &      &      &       & $\times$10$^{15}$ [cm$^{-2}$] \\ 
   \midrule
    119 & FTH150P & R4S50x50-P1  &  p & 2.4 \\  
    120 & FTH150P & R4S100x25-P1 &  p & 2.4 \\
    166 & FTH150P & R4S50x50-P8  &  p & 2.3 \\           
    174 & FTH150P & R4S100x25-P1 &  p & 5.4 \\
    176 & FTH150P & R4S50x50-P8  &  - & 0.0 \\       
    179 & FTH150P & R4S100x25-P7 &  p & 5.4 \\
   191 & FTH150P & R4S50x50-P2  &  p & 2.3 \\               
    193 & FTH150P & R4S100x25-P7 &  p & 5.4 \\                          
    194 & FDB150P & R4S100x25-P1 &  n  & 3.6 \\
    195 & FDB150P & R4S100x25-P1 &  n  & 0.5 \\    
    196 & FDB150P & R4S100x25-P1 &  n  & 7.2 \\        
   197 & FDB150P & R4S100x25-P7 &  n  & 14.4 \\            
   198 & FDB150P & R4S50x50-P1  &  n  & 7.2 \\
    202 & FTH150Y & R4S50x50-Y2  &  n  & 3.6 \\      
    509 & FTH150P & RD53A100x25-P1 &  p & 5.2 \\                
    512 & FTH150P & RD53A100x25-P2 &  p & 5.6 \\            
  \bottomrule
  \end{tabular}
  \label{tab:Samples}
\end{table}

\end{document}